\documentclass[twocolumn]{aa}
\usepackage{graphicx}
\usepackage{txfonts}
\def\D{{\rm d}}

\begin{document}

\title{Type Ia Supernova models arising from different distributions of igniting points}

\author{D. Garc\'\i a-Senz\inst{1,2} \and E. Bravo\inst{1,2}} 

\offprints{D. Garc\'\i a-Senz}

\institute{Departament de F\'\i sica i Enginyeria Nuclear, UPC, Sor 
Eulalia d'Anzizu s/n, B5, 08034 Barcelona, Spain\\ 
email: domingo.garcia@upc.es, eduardo.bravo@upc.es 
\and 
Institut d'Estudis Espacials de Catalunya} 

\date{Received .....; accepted ???}

\abstract{
In this paper we address the theory of Type Ia supernovae from the moment of
carbon runaway up to several hours after the explosion. We have concentrated
on the boiling-pot model: a deflagration characterized by the (nearly-)      simultaneous ignition of
a number of bubbles that pervade the core of the white dwarf. Thermal
fluctuations larger than $\ga1$\% of the background temperature 
($\sim7\times10^8$~K) on lengthscales of $\le 1$~m could be  
the seeds of the bubbles. Variations of the homogeneity of 
the temperature perturbations can lead to two
alternative configurations at carbon runaway: if the thermal gradient is small,
all the bubbles grow to a common characteristic size related to the value of 
the thermal gradient, but if the thermal gradient is large enough, the size spectrum of the bubbles extends over several orders of magnitude. The explosion
phase has been studied with the aid of a smoothed particle hydrodynamics code
suited to simulate thermonuclear supernovae. In spite of important
procedural differences and different physical assumptions, our results converge
with the most recent calculations of three-dimensional deflagrations in white
dwarfs carried out in supernova studies by different groups. For large initial numbers of
bubbles ($\ga3-4$ per octant), the explosion produces $\sim0.45 M_{\odot}$ of
$^{56}$Ni, and the kinetic energy of the ejecta is $\sim0.45\times10^{51}$~ergs.
However, all three-dimensional deflagration models share three main drawbacks:
1) the scarce synthesis of intermediate-mass elements, 2) the loss of 
chemical stratification of the ejecta due to mixing by Rayleigh-Taylor
instabilities during the first second of the explosion, and 3) the presence of
big clumps of $^{56}$Ni at the photosphere at the time of maximum brightness. On
the other hand, if the initial number of igniting bubbles is small enough, the
explosion fails, the white dwarf oscillates, and a new opportunity comes for
a detonation to ignite and process the infalling matter after the first
pulsation.

\keywords{Supernovae: general, - hydrodynamics, nuclear reactions, nucleosynthesis}  
}
\titlerunning{Type Ia Supernovae models arising from .....} 
\maketitle

\section{Introduction}

The huge increase in number, quality and diversity of observational data 
related to Type Ia supernovae (SNIa) in recent years, 
combined with the advance in computer technology, have persuaded modellers
to leave the phenomenological calculations that rely on   
spherical symmetry, and attempt more physically meaningful 
multidimensional simulations. Although the more plausible models of the 
explosion
always involve the thermonuclear disruption of a white dwarf, the current zoo 
of explosion mechanisms is still
too large to be useful in cosmological applications of Type Ia supernovae
 or to make it possible to understand the details of the chemical evolution
of the Galaxy. Nowadays, the favoured SNIa
model is the thermonuclear explosion of a white dwarf that approaches the 
Chandrasekhar-mass limit owing to accretion from a companion star at the 
appropiate rate to avoid the nova instability. 
Other types of models, such as the sub-Chandrasekhar models or the
double degenerate scenario, although not completely ruled out, either are not
able to explain even the  
gross features of the spectrum and light curve of normal  
supernovae, or face   
theoretical objections (Nugent et al.~\cite{n97}, Napiwotzki et
al.~\cite{n02}, Segretain, Chabrier \& Mochkovitch~\cite{scm97}, Saio \& 
Nomoto~\cite{sn98}). 

Admitting that at least part of the diversity shown by the Type Ia supernovae
sample
is directly related to the explosion mechanism of a Chandrasekhar-mass 
white dwarf (Hatano et al. \cite{hblbfg00}), current multidimensional 
hydrodynamical calculations should be able to 
allow for a wide range of nickel masses synthesized in different 
events and, at the same time, to reproduce the stratified chemical profile
suggested by SNIa observations (e.g. Badenes~\cite{b04} and references 
therein). The few
multidimensional calculations carried out so far
(Reinecke, Hillebrandt \& Niemeyer \cite{rhn02}, RHN hereafter, Gamezo  et al. 
\cite{gkochr03}, G03 hereafter, and Bravo \& Garc\'\i a-Senz \cite{bgs03}, for 
the most recent results) have shown interesting 
deviations from what is predicted in spherically symmetric models: a) the   
geometry of the burning front is no longer spherical owing to the important 
role played by buoyancy and hydrodynamical instabilities, b) the explosion is 
stronger when calculated in three dimensions (3D) than it is in two-dimensional 
simulations (2D), c) there is an 
inhomogeneous chemical structure,  
in particular the amount of $^{56}$Ni is sufficient to power the light curve, 
but it is localized 
in pockets distributed all along 
the radius of the white dwarf and, d) an uncomfortable amount of carbon and 
oxygen remains unburnt, most of it close to the outer layers but also at the
very center of the 
white dwarf. Lacking multidimensional studies of the light curve and spectra, 
it is certainly risky to interpret the results of the aforementioned  
3D-hydrosimulations in order to elucidate the explosion mechanism and 
discriminate between models. Nevertheless, one can already wonder if there is 
any significant observational evidence of departure from spherical symmetry, and 
to what extent the presence of mixed material and unburnt carbon might be 
detectable in the spectrum of Type Ia supernovae. In this regard, there is a 
number of indications that the departure from spherical symmetry is not 
large. These include the low level of polarization in most SNIa, 
although there are exceptions (see, for instance, Kasen et al. \cite{knw03}), 
the quite homogeneous profile of 
the absortion line of SiII, which points towards a limited amount of clumping in
the ejecta (Thomas et al. \cite{tkbb02}), and the fact that galactic and
extra-galactic supernova 
remnants (SNR) do not show large departures from spherical symmetry (i.e. the 
blast wave of Tycho's SNR). 
Regarding the 
chemical composition of the ejected matter, recent spectroscopic observations
of a dozen  {\sl Branch-normal} Type Ia supernovae in the near infrared (Marion 
et al. \cite{mhvw03}) suggest 
that the unburnt matter ejected has to be less than $10\%$~of the mass of the 
progenitor. According to these results, the presence of a substantial 
amount of unburnt low-velocity 
carbon near the center of the star is rather improbable (but see also  
Baron, Lentz \& Hauschildt \cite{blh03}).

The 3D calculations of pure deflagrations presented in
RHN and G03 were based on different 
hydrodynamical codes and algorithms to track the propagation of the nuclear 
flame, and started from different initial conditions. In spite of this, both 
calculations reached similar conclusions: the mass of radioactive nickel
synthezised in the explosion was enough to power the light curve, the 
kinetic energy of the ejecta was of the order of $10^{51}$~erg, a small amount
of intermediate-mass elements was 
synthesized, and the chemical composition was radially mixed and spread in 
velocity space.

In this paper we present a set of three-dimensional simulations of 
the thermonuclear explosion of a white dwarf carried out with an SPH code. The 
initial conditions were similar to those in RHN, i.e.   
a number of bubbles of burnt material scattered around the 
center of the white dwarf. Our objective was to explore the dependence of the
results on the initial configuration of the igniting bubbles. We 
reproduced the main results of RHN for similar initial conditions. With the
confidence that this convergence gave us, we explored thereafter different
configurations.
Special care was taken in the setting of the initial conditions of the 
explosion and in the characterization of the degree of clumping of the 
different elements present in the ejecta. As we explain below, we obtained 
explosions 
within the correct range of kinetic energy, $^{56}$Ni mass, and isotopic
composition of the Fe-peak  
elements, provided the number of  
hot spots is larger than a critical value. However, in all the simulations a  
significant amount of unburnt fuel was ejected, clumped in pockets scattered all 
over.
Below a critical number 
of hot spots the white dwarf remained bound, 
and a phase of recontraction ensued which could lead either to an off-center
carbon detonation 
(the 3D analog of the pulsating delayed detonation), or even to 
the detonation of the undesirable carbon left around the center
(Bravo \& Garc\'\i a-Senz 2005, in preparation). 

The organization of the text is as follows. In 
Sect. 2 we try to shed some light on the difficult problem of the initial 
conditions of the explosion  using analytical means, but also by 
building  a toy 
numerical model. Section 3 is devoted to summarizing the relevant features of the
hydrocode and to providing the main results of the simulations. The evolution of 
the ejecta up to several hours after central ignition is discussed in Sect. 4.
Some discussion and the main conclusions of our 
work as well as the prospects for future work are provided in Sect. 5.
Finally, we develop in an Appendix a novel statistical treatment of the initial 
conditions at the time of thermal runaway.

\section{The early stages of the explosion} 

There is growing evidence that the outcome of the thermonuclear explosion
of a white
dwarf relies on the poorly understood stage which precedes carbon ignition. 
A few minutes prior to the runaway, the central part of the white dwarf 
is in a highly turbulent state and the path to the explosion of a fluid element
is determined by the interplay between heating due to the collision of 
turbulent eddies and nuclear energy generation, and cooling due to the 
expansion of the fluid element and electron conduction. In order to know when and 
where the carbon runaway begins and which is the geometry of the 
ignited region, it would be necessary to perform calculations of this phase 
in three dimensions with a good resolution over a significant period of time 
(compared to the Courant time). The problem is so complex that direct 
simulations are not feasible and the only way to gain insight  
into the conditions at runaway is by combining analytical ideas and toy 
models with 
one and  
two-dimensional simulations. 

H\"oflich \& Stein (\cite{hs02}) carried out the first (and, as far as we know,
the only one published insofar) 
two-dimensional calculation of the 
evolution of the white dwarf prior to the carbon runaway using an implicit
hydrocode. Their simulation strongly suggests that the ignition takes place 
in one or a few localized spots, referred to {\sl bubbles} or {\sl blobs} from  here on, amidst a
stirred environment. According to 
their results the convective elements have a characteristic velocity of about 
100~km~s$^{-1}$, and carry a kinetic energy 
$\simeq5~10^{13}$~ergs~g$^{-1}$. Assuming that 
collisions between the convective elements can efficiently convert their 
kinetic energy into thermal energy, this 
results in   
temperature fluctuations $\Delta T/T_\mathrm{bg}\simeq 0.01$~(for a background temperature
$T_\mathrm{bg}=5~10^8$~K).
Given  adequate physical conditions, these fluctuations could grow to   
finally trigger the nuclear runaway. If the thermal fluctuations encompass a 
large enough initial volume, so that heat conduction can be neglected, their 
evolution is governed by the balance between nuclear 
heating and cooling by expansion. In this case, the buoyancy of the fluid 
elements opens the possibility of  
having off-center ignition, provided the nuclear characteristic timescale  
becomes similar to the expansion cooling timescale of the rising bubbles
(Garc\'\i a-Senz \& Woosley \cite{gsw95}). 

Things are different for very localized 
fluctuations. There, cooling by electron conduction  
through the bubble surface becomes relevant, and the fate of any fluctuation depends on 
the balance between  release of nuclear energy and the combined heat losses by conduction and small-scale turbulence caused by convection. Nevertheless, lacking a quantitative model for the convection the impact of small-scale turbulence is difficult to estimate. An approximate value for the Nusselt number, which
gives the ratio between the total rate of heat transfer and 
conduction over the length-scale $L$   
for typical conditions at the ignition stage was provided by Woosley et al.  (2004), Nu$\simeq 3~10^{11}$~for $L=200$~ Km. Assuming a
Kolmogorov-like model for turbulence the resulting Nu number at scale
$\lambda$ is Nu$(\lambda)=\mathrm {Nu}(L)(\lambda/L)^{4/3}$. Thus, heat surface losses by
conduction become dominant for Nu$(\lambda)\leq 1$, on millimetric 
length-scales 
. However, given the very qualitative nature of the Kolmogorov 
scaling law and other 
uncertainties, this critical length-scale could be larger, up to several
centimetres or even more. 

According to the above discussion  a practical condition for  
ignition can be set especially suited for small enough regions where conduction takes
over.  
Let us suppose a Gaussian fluctuation of the 
background temperature $T_\mathrm{bg}$:

\begin{equation}
T=T_\mathrm{bg}+\Delta T \exp{\left(-\frac{(x-x_0)^2}{\delta^2}\right)}
\end{equation}

\noindent
where $\delta$~is the characteristic spatial size of the
fluctuation and
$x_0$~is the center of the bubble.  Neglecting the effect of small-scale 
turbulence a 
necessary (although, in general, not sufficient) condition for the fluctuation to grow
can be obtained by equating the conductive cooling to
the nuclear energy generation rate in the 
center of the bubble. This leads to the following expression which 
relates the spatial size of the spot to the temperature fluctuation:

\begin{equation}
\delta=\sqrt{\frac{2\sigma\Delta T}{\rho~\dot S_\mathrm{nuc}}}
\end{equation}

\noindent 
where $\sigma$~is the thermal conductivity and $\dot S_\mathrm{nuc}$~is the 
nuclear energy generation rate.
The resulting ignition lines for $\rho_9=2$~and several background temperatures are depicted 
in Fig. 1. Sucessful ignition is only possible above the corresponding
line,  
otherwise conductive losses are able to smear the thermal fluctuation. As 
can be seen, larger background temperatures may lead to 
the carbon runaway even for fluctuations of spatial size lesser than one
meter, whereas smaller environmental temperatures demand fluctuations exceeding
several meters. We have studied the evolution to runaway of one of these 
bubbles by solving the diffussion equation in the planar approximation jointly  
to a nuclear network implicitly coupled with temperature (Cabez\'on, Garc\'\i 
a-Senz \& Bravo \cite{cgsb04}) and using a very fine zoning. The initial 
thermal profile was that given in Eq. (1) with $T_\mathrm{bg}=7.5~10^8$~K and $\Delta 
T/T_\mathrm{bg}=0.01$. The thermal evolution of the bubble is shown in Fig. 2, where it
can be seen that the elapsed time to ignition is lesser than 3 seconds.  This 
time is similar to the time of residence of the convective elements in the 
core (Woosley et al. 2004). Thus, the evolution shown in Fig. 2 is only 
approximate, as convection has not been taken into account. During 
that time, the radius of the bubble grows from its initial value
$R_\mathrm{b} (t=0~\mathrm{s})\simeq 20$~cm 
to $R_\mathrm{b} (t=2.9~\mathrm{s})\simeq 40$~cm while the central temperature 
climbs to $10^9$~K and carbon burning ensues. Once a nuclear flame is born 
in these tiny regions it takes less than 0.1 seconds to conductively propagate 
the combustion to a few kilometers (Timmes \& Woosley \cite{tw92}). At 
this point, the Archimedean force becomes high enough to accelerate the bubble
to a substantial fraction of the local sound speed. In this way, the nuclear 
flame can rapidly be transported out of the core of the white dwarf, marking 
the beginning of the thermonuclear explosion. But, during the few seconds elapsed 
in the making of the flame, there is a non-negligible chance for other regions
of the core to develop localized runaways. On the whole, the white dwarf core 
would resemble a {\sl boiling fluid} in which initially small bubbles with a 
temperature slightly higher than their surroundings grow in size (as 
they are fed by nuclear combustion), to finally float away when their 
radius becomes larger than a critical size, of the order of one 
kilometer.  

\begin{figure}
\resizebox{\hsize}{!}{\includegraphics{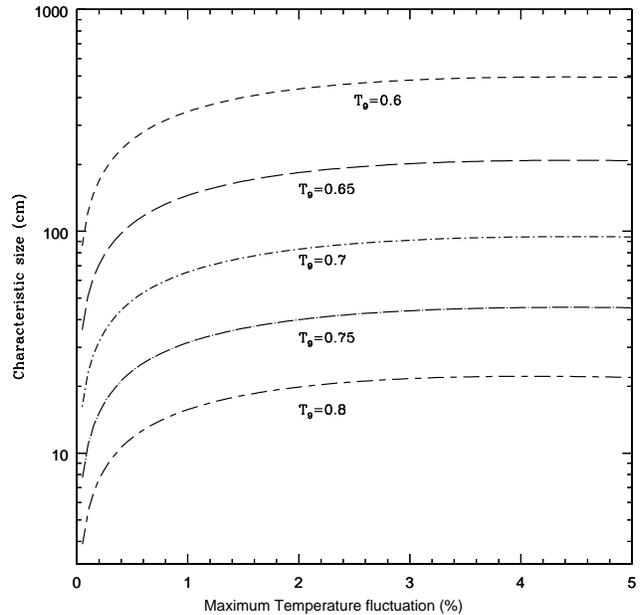}}
\caption{Ignition lines giving a necessary condition for a thermal
fluctuation to grow and develop a carbon runaway in a localized region. 
Horizontal axis is the size of the fluctuation, $\Delta T/T_\mathrm{bg}$~ at the center
of the bubble (see Equation 1). 
Vertical axis is the radius encompassed by the 
fluctuation and each one of the five lines represent a particular background
temperature $T_\mathrm{bg}$. The area above the corresponding line gives the locus of 
successful ignitions. 
}\label{fig1}
\end{figure}

\begin{figure}
\resizebox{\hsize}{!}{\includegraphics{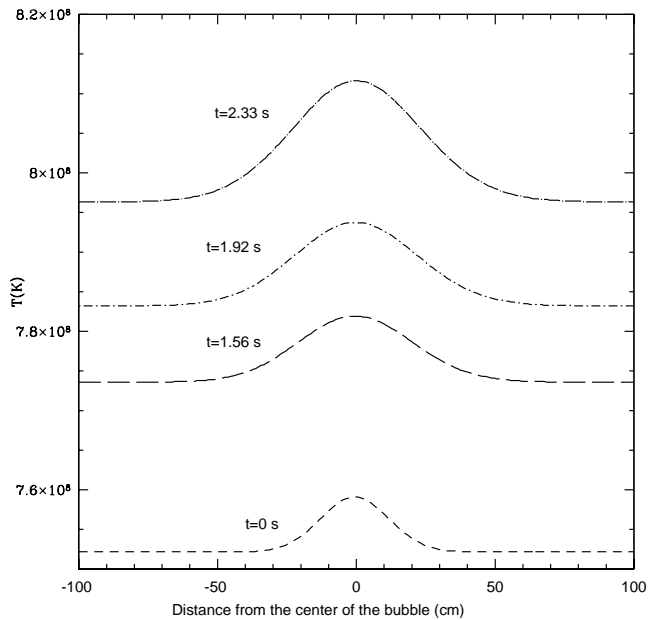}}
\caption{Evolution of the thermal profile within the bubble from the initial
fluctuation until the carbon runaway. It takes less than 3 seconds for the 
center of the bubble to reach $10^9$~K. During this time the bubble 
radius increases by a factor of two and the background temperature rises owing 
to the release of nuclear energy.  
}\label{fig2}
\end{figure}

\subsection{Sizes and distribution of the bubbles}

We have 
developed a statistical model for the igniting 
bubbles, whose details are explained in the Appendix, and whose main results are 
given here. We start by considering an initial distribution of 
hot spots, characterized by their peak temperatures, $T_0$, and their thermal 
profile. The free parameters of our model are the initial distribution function
of peak temperatures, $\theta\left(T_0\right)$, and the thermal profile of any bubble at
the time its center reaches the ignition temperature, $10^9$~K. The 
characteristic lengthscale of the thermal profile of each igniting bubble will 
be denoted by $R_0$. Our target is to determine the
distribution function of the sizes (radii) of the igniting bubbles as a function of
 time: $\D N/\D R_\mathrm{b}$.

In the course of time each hot spot ignites at its center and begins growing
due to the combined effect of spontaneous combustion and conductive flame
propagation. We have been able to calculate the radius of a given bubble, with a
given initial peak temperature, as a function of time (see Appendix for 
details). Afterwards, we computed
the distribution function of the radii of the
bubbles, $R_\mathrm{b}$, as a function of time for different sets of functional
dependences of $\theta\left(T_0\right)$, and different values of $R_0$ (Eqs. A.8
and A.9). 

We have found that the results are insensitive to the functional form of
$\theta\left(T_0\right)$, within reasonable choices. In contrast, the
lengthscale of the thermal profile, $R_0$, is the most influential parameter for 
the temporal evolution of the size of the bubbles. Basically, our results show
that, depending on the initial thermal gradient inside each hot spot, the
bubbles
can follow two different regimes.
If the thermal profile is shallow enough, they grow
due to spontaneous flame propagation, otherwise they
grow conductively. In fact, the process of conductive growth of the bubbles is
always preceded by a phase of spontaneous propagation. 

\begin{figure}
\resizebox{\hsize}{!}{\includegraphics{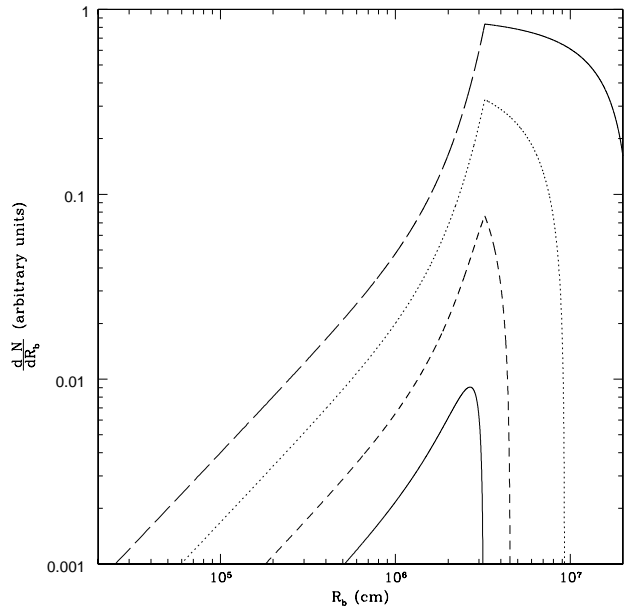}}
\caption{Evolution of the distribution function of radius of the igniting 
bubbles as
a function of time since ignition, for $R_0 = 10^7$~cm. The lines correspond to 
times $t = 0.1$, 0.3, 1, and 3~s.  
}\label{fig3}
\end{figure}

 Fig.~\ref{fig3}  shows the temporal evolution of the distribution
function of the size of the bubble for a particular value of $R_0$. Other choices of
$R_0$ lead to the same kind of temporal evolution, although with different
scales of time and length. Initially, the distribution is driven by the 
spontaneous ignition of matter. The bubbles grow very fast at the beginning,
because of the flat thermal gradient at their centers, but the phase velocity 
soon
drops due to the decreasing temperature found at increasing distances to the
center of the bubble. The result is a distribution function with a pronounced peak at a
characteristic radius,  and with a shape which recalls an impulse function. 
With time the phase velocity drops below the conductive flame velocity, and
the shape of the distribution function changes, flattening above the transition 
radius. Actually, the distribution corresponding to the late time shown in 
Fig.~\ref{fig3} would never be
realised, because of the bubbles' tendency to float to lower density regions.

\begin{figure}
\resizebox{\hsize}{!}{\includegraphics{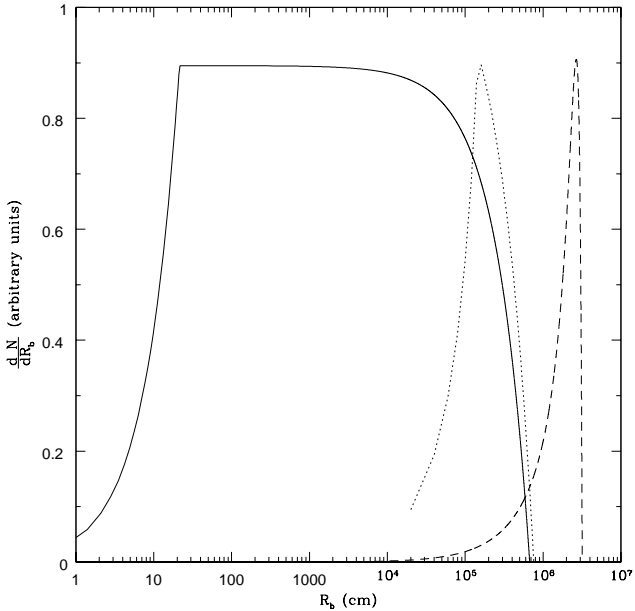}}
\caption{Distribution function of the radii of the igniting bubbles at
a time $t = 0.1$~s, for different values of $R_0$: $10^4$~cm ({\sl solid line}),
$10^6$~cm ({\sl dotted line}), and $10^7$~cm ({\sl dashed line}). Note the
linear scaling of the vertical axis in this figure, in contrast with the
logarithmic scaling of the previous one.
}\label{fig4}
\end{figure}

The different kinds of solutions found for different values of 
$R_0$ can be seen in Fig.~\ref{fig4}, where the distribution function of the 
sizes 
of the bubbles is shown at time $t = 0.1$~s, and for values of $R_0 = 10^4$, 
$10^6$, and $10^7$~cm. For the two largest values of $R_0$, the configuration
consists of an arbitrary number of bubbles of nearly {\sl equal size}. This 
size is of the order of the transition radius from 
spontaneous to conductive propagation, $R_\mathrm{tr}$ (see Fig.~\ref{figa1}). 
In contrast, if the value of $R_0$ is small enough (which
means that the thermal gradient at the moment of central ignition in each hot
spot is steep enough) bubbles of different radii are produced. In this  
case the distribution function achieves a nearly constant value through 
several orders of magnitude in $R_{\mathrm b}$. This results in the coexistence 
of bubbles of quite different dimensions. In any case, the statistical 
approach
becomes justified in view of the ratio between the total volume of the 
adiabatic core, which extends up to a radius of $\sim400$~km, and the  
volume of a typical bubble at $t = 0.1$~s, the volume ratios being $\sim2000$, 
$\sim2\times10^7$, and 
$\sim10^{12}$, 
 for $R_0 = 100$, $10$, and $0.1$~km, respectively. On another note,
it could be hard to achieve a smooth thermal gradient over large distances owing to
the highly turbulent state of the core at this stage of the evolution of the 
white dwarf. Therefore, we believe that
a continuum distribution of radii of the blobs (as that represented by the continuum
line in Fig.~\ref{fig4}) might be easier to realize.

It is illustrative to calculate the total area of the bubbles when the distribution 
function becomes independent of $R_{\mathrm b}$ in a range of radii between
$R_\mathrm{min}$ and $R_\mathrm{max}$, which is more or less the case  
represented by the solid line in Fig.~\ref{fig4}: 
\begin{equation}
A = \int_{R_\mathrm{min}}^{R_\mathrm{max}}{4\pi R_{\mathrm b}^2 
\frac{\D N}{\D R_{\mathrm b}}\D R_{\mathrm b}} \simeq 
\frac{4}{3}\pi R_\mathrm{max}^3 \frac{\D N}{\D R_{\mathrm b}}.
\end{equation}
\noindent
Calculating in the same way the total
volume, $V$, occupied by the bubbles gives $A/V = 4/R_\mathrm{max}$. 

\section{Three-dimensional hydrodynamical models}

In this work, we have focused on the study of the first kind of initial 
configuration found in the previous section, i.e. that formed by an
arbitrary number of bubbles of {\sl equal size}. Even though a configuration of
bubbles of different sizes seems more probable, its simulation is more
computationally demanding. Thus, we have computed 3D models of 
explosions
starting from varying numbers of bubbles, and we have tested the dependence of
the results on the number of bubbles. We have also computed a
model whose initial configuration is made up bubbles of different sizes, to obtain 
a first insight in the kind of variations of the results introduced by the
second type of configuration found in the previous section. Further tests of
this kind of configuration are deferred to  
future work.

The setup of the initial conditions of a 3D hydrodynamical simulation of a
thermonuclear explosion requires choosing several parameters. In
principle, it would be nice to test the dependence of the simulation results on
all these configurational parameters, but in practice this would demand an
unaffordable
computational effort. However, some insight can be gained by the analysis of the
dependence of the area-to-volume ratio of different initial configurations. 
Consider,
for instance, the case addressed in Eq.~3 and in the last paragraph of the
previous section (let us call it the multi-bubble configuration). One can 
compare these results with those corresponding to a 
single spherical bubble. If one chooses the single bubble to have the same
radius as the maximum radius of the multi-bubble configuration,
$R_\mathrm{max}$, then the area-to-volume ratio is larger in the multi-bubble
case by a factor 4/3. Hence, one can expect an
increase in the flame velocity (or, more precisely, in fuel consumption rate) 
of the order of 33\%. On the other hand, if the comparison
is with respect to a single bubble occupying the same volume as a
configuration of N bubbles (thus, having the same mass in both cases), 
then the area-to-volume ratio is larger in the multi-bubble case by a factor:
\begin{equation}
\frac{A}{A_\mathrm{s}} = \frac{1}{3}\left(16R_\mathrm{max}
\frac{\D N}{\D R_{\mathrm b}}\right)^{1/3} \simeq 0.84 N^{1/3}, 
\end{equation}
\noindent
so, at least at the beginning, there is a clear dependence of the fuel 
consumption rate on the initial number of bubbles.

All the 3D models were calculated using a SPH code with 250,000 particles of 
identical mass. The main features of our hydrocode
can be found in Garc\'\i a-Senz, Bravo \& Serichol (\cite{gsbs98}). 
The energy transport from the hot burnt matter to the fresh fuel was 
simulated by re-scaling the conductivity and the nuclear energy generation 
rate in the diffusion equation in such a way that the flame moved  
with a prescribed constant velocity, $v_\mathrm{b}$ 
(Garc\'\i a-Senz et al. \cite{gsbs98}). 
The nuclear network used in the hydrocode consists of a 9 nuclei chain 
(Timmes, Hoffman \& Woosley 
\cite{thw00}), which is basically an $\alpha$-network until $^{32}$S plus a 
direct link to $^{56}$Ni. When the temperature became higher than $5\times10^9$~K 
 nuclear statistical equilibrium (NSE) was assumed, and the nuclear binding 
energy, electron capture rate, and the molar fractions of nuclear species were 
interpolated from a table. Detailed 
nucleosynthesis was calculated by postprocessing the output of the 
hydrodynamics of the most relevant models. The equation of 
state has contributions from relativistic electrons of arbitrary degeneracy
(with pair corrections), an ideal gas of ions with electrostatic 
interactions, and radiation.

\subsection{Initial setup}

The initial configuration of the computed models is given in Table~\ref{tab1}.
There $N_\mathrm{b}$ is the initial number of bubbles in the model. Each
initial model for the SPH calculation was built in four steps. After the
first two steps, the mechanical structure of the white dwarf was
reproduced by generating a hydrostatically stable particle
distribution.  In the last two steps, the thermal and chemical profile
representative of the bubbles was  imprinted on the model, while maintaining
the particles at rest. The values of the radii of the bubbles, $R_\mathrm{b}$, 
and total incinerated mass, $M_\mathrm{b}$, given in Table~\ref{tab1}
refer to the end of the third and fourth steps, respectively.

\begin{table}
\centering
\caption[]{Models and initial characteristics of the bubbles.}
\label{tab1}
\begin{tabular}{lcrcc}
\hline
\noalign{\smallskip}
Model & $\rho_\mathrm{c}$ & $N_\mathrm{b}$ & $R_\mathrm{b}^\mathrm{a}$ &
$M_\mathrm{b}/M_\mathrm{WD}^\mathrm{a}$ \\
 & (g cm$^{-3}$) & & (km) \\
\noalign{\smallskip}
\hline
\noalign{\smallskip}
B30U  & $1.8\times10^9$ & 30 & 43-60 & 2.8\% \\
B10U  & $1.8\times10^9$ & 10 & 63-89 & 3.0\% \\
B07U  & $1.8\times10^9$ &  7 & 49-60 & 2.8\% \\
B06U  & $1.8\times10^9$ &  6 & 49-60 & 2.7\% \\
B90R & $1.8\times10^9$ & 90 & 11-92 & 2.9\% \\
\noalign{\smallskip}
\hline
\end{tabular}
\begin{list}{}{}
\item[$^{\mathrm{a}}$] Sizes and masses of the bubbles are given here at different 
times. See text for an explanation.
\end{list}
\end{table}

The mechanical structure of the initial SPH model was obtained by relaxing 
a three-dimensional sample of particles, radially distributed to match the 
mass distribution of a white dwarf of $1.38 M_\odot$. The thermal profile of 
the white dwarf was taken to be adiabatic in the central region, and 
isothermal beyond a radius of 400~km.
The relaxation process was divided into two
steps: a) no radial displacement of any particle was allowed  whereas a
dissipative force, proportional to the tangential velocity, was added to the
equations of motion, in order to approach a spherically symmetric distribution
of mass points and, b) the dissipative force was removed and  radial
displacement allowed, in order to  reach the final stable distribution.   
The relaxation ended once both the density profile and the radial pressure 
gradient matched the one-dimensional white dwarf structure. 
The nominal resolution of the SPH calculations is given by the
smoothing length, $h$\footnote{Those unfamiliar with SPH codes should note that
the smoothing length does not give the separation between the mass particles, 
but its value limits the resolution of the simulation in the sense that features
smaller than $h$ are smoothed out. In the present simulations, we worked with a
mean of 50 neighbors per particle, implying that about 50 particles were present
inside a sphere of radius $2h$}, which achieved a minimum value of 
$\simeq20$~Km in the
central zones and scaled outwards as $(\rho_c/\rho)^{1/3}$. 

The third step in the generation of the initial model was the assignment of
particles to the bubbles. First, once  the sizes of the bubbles were decided, their 
position was randomly chosen within
the central 400~km of the white dwarf, with the only constraint that the bubbles
should
not overlap. Next, the SPH particles which were located inside a bubble were
marked as incinerated particles, their temperature was raised isochorically to
NSE equilibrium, and their chemical composition and nuclear binding energy was
changed accordingly. This procedure produced a random realization of the
bubbles, which resulted in bubbles of radii $R_\mathrm{b}$ as given in
Table~\ref{tab1}. The first four models in this table 
correspond to bubbles of (nearly) the same size, while the last one was devised to
reproduce a distribution of bubbles according to the law: $\D N/\D R_{\mathrm b} \propto R_{\mathrm
b}^{-0.8}$. The lower end of the size spectrum  actually covered by the last
model was strongly limited by the resolution of the code, as can be appreciated
in the table. 

The fourth and last step consisted of the hydrostatic growth of the bubbles
through thermal conduction and nuclear reactions. This step is necessary  
 to generate a well defined diffussive flame structure around each bubble,
before starting the simulation of the supernova explosion. At the end of this 
step, the
mass burnt in all the bubbles was $M_\mathrm{b}$ as given in
Table~\ref{tab1}. As can be seen, the mass burned is almost the same in all
the computed models, allowing  a meaningful comparison between them. 

\begin{figure*}
\centering
\resizebox{\hsize}{!}{\includegraphics{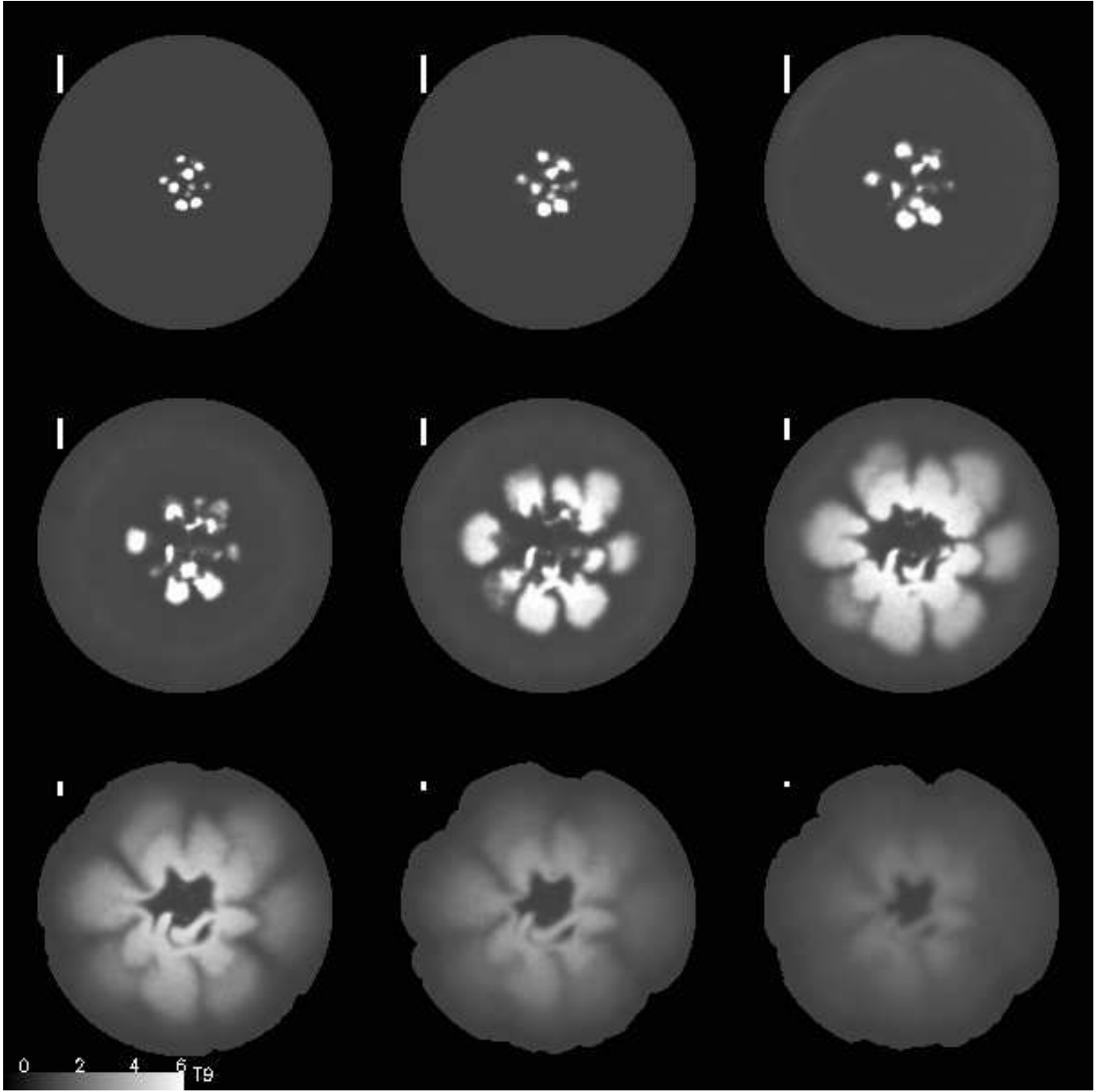}}
\caption{Snapshots of the temperature distribution in a meridian plane of
model B30U at times from 0 to 1.3 s, in steps of $\sim0.15$~s. The 
temperature scale is shown at the bottom left of the image, while
the length scale is shown at the top left of each snapshot (the
length of the vertical bar is equivalent to 200~km).
}\label{fig5} 
\end{figure*}

\subsection{Comments on the numerical consistency of the calculations}

The numerical noise present in the initial
model must be as low as possible to not interfere with the rising of the  
hot elements, especially at the beginning, when they move very subsonically. 
For given initial conditions (i.e. number and size of ignited blobs) the 
evolution of the model is determined by the interplay between the Archimedean 
force and the effective combustion through the bubble surface. While the 
correct buoyancy can be reasonably reproduced by taking large enough bubbles
and 
minimizing the numerical viscosity of the code, the effective combustion rate 
is hard to treat numerically. During their rising, the incinerated chunks of 
material 
are subjected to the Kelvin-Helmholtz and other hydrodynamical instabilities 
which greatly deform their geometry. Such deformation takes place in a
range of scalelengths which can not be fully resolved by any present hydrocode 
neither in 3D nor in 2D. 
On the other hand, the increase of the bubble area competes with 
surface destruction due to the collision between the floating elements. It has
been argued that these two antagonic effects give rise to a
self-regulating mechanism which finally decouples the macroscopic burning 
velocity from the microphysics. Even though there are some indications that
such a self-regulating mechanism could be at work during the development of the 
explosion (Khokhlov \cite{k95}, Gamezo et al. \cite{gkochr03}, 
Garc\'\i a-Senz \& Bravo \cite{gsb03}) 
its existence has not yet been proved. In this work, we have adopted a 
practical 
point of view by incorporating the effect of those scalelengths not resolved 
by the SPH simulation through a constant effective burning velocity that we called the 
baseline
velocity, $v_\mathrm{b}$. 

In the present models, we adopted a value of 
$v_\mathrm{b} = 200$~km~s$^{-1}$, which is similar to what can be found in
current supernova literature. Due to the
self-adjustment of the flame surface, the dependence of  the outcome  of
the explosion on the particular value adopted for $v_\mathrm{b}$ is much 
weaker than that observed in one-dimensional models with respect to the
parametrized burning front velocity. We have  recalculated model B30U with a
baseline velocity of 100~km~s$^{-1}$ and have obtained variations in the
released nuclear energy and $^{56}$Ni mass of only 8\%. Paradoxically, it was 
the smallest flame velocity
run which burned most fuel and released most nuclear energy.

Of particular importance is the amount of  
spureous viscosity introduced by the hydrocode, because it could prevent the
rise of  hot blobs below a critical size. For example, for a small sphere 
subjected to an effective gravity, $g_\mathrm{eff}=g(\Delta\rho/\rho)$, the 
outward 
acceleration is given by: 
$a(r) = g_\mathrm{eff}(r) - F_\mathrm{D}(r)/m_\mathrm{b}(r) - 
F_\mathrm{S}(r)/m_\mathrm{b}(r)$, where $m_\mathrm{b}$ is the mass of the 
sphere, $F_\mathrm{D}$ is the drag force and $F_\mathrm{S}$ is the
Stokes force induced by shear viscosity. A common expression used to 
calculate the drag force is $F_\mathrm{D} = 
0.5 C_\mathrm{D}\rho v^2\pi R_\mathrm{b}^2$ (where $C_\mathrm{D}$ is the 
drag coefficient, of order unity). For an ascending sphere the 
Stokes force is given by $F_\mathrm{S}=6\pi\rho\nu R_\mathrm{b} v$, where 
$\nu$~is the kinematic 
viscosity coefficient and $v$ is the ascending velocity of the sphere relative
to the surrounding fluid. Therefore $F_\mathrm{D}/m_\mathrm{b}\propto 
v^2/R_\mathrm{b}$~and $F_\mathrm{S}/m_\mathrm{b}\propto \nu v/R_\mathrm{b}^2$. 
In a white dwarf $\nu$~is very low and the Stokes 
force completely negligible. Nevertheless, the numerical viscosity introduced 
by any hydrocode is many orders of magnitude larger than the actual microscopic
viscosity. Given the different dependence on $R_\mathrm{b}$ and $v$ shown in 
the previous expressions for
$F_\mathrm{D}/m_\mathrm{b}$ and $F_\mathrm{S}/m_\mathrm{b}$, the numerical
viscosity can introduce a non-negligible force during the first
stages of the ignition phase, when the radius and the velocity of the burned 
blobs still remain small. For large 
enough blobs, the Stokes force rapidly becomes weaker than the drag force, and the 
numerical viscosity is not expected to interfere with the simulations. In SPH, one source of 
viscosity (although not the only one) is the artificial viscosity term devised to 
handle shock waves. Roughly, this term introduces  an amount of viscosity 
$\nu\simeq \alpha h c_\mathrm{s}+\beta h v$ where $\alpha\simeq\beta\simeq 1$ 
are two 
parameters related to the linear and the quadratic terms present in the standard
artificial viscosity formulation, and $c_\mathrm{s}$ is the local sound speed. 
Unfortunately, one can not choose $\alpha=\beta=0$ without compromising the 
stability of the initial model. We found that a reasonable choice is to take 
$\alpha=0,~\beta=1$, which, for our initial configuration of bubbles, leads to 
a Reynolds number of about five. As explained before, the procedure to set the initial conditions of 
our calculations ends with a brief hydrostatic episode of burning and 
heat conduction, until the flame around the blob is built. This 
strategy allows an 
increase in the size of the blobs, thus reducing the damping effect of the 
spurious numerical viscosity. 

\begin{figure}
\resizebox{\hsize}{!}{\includegraphics{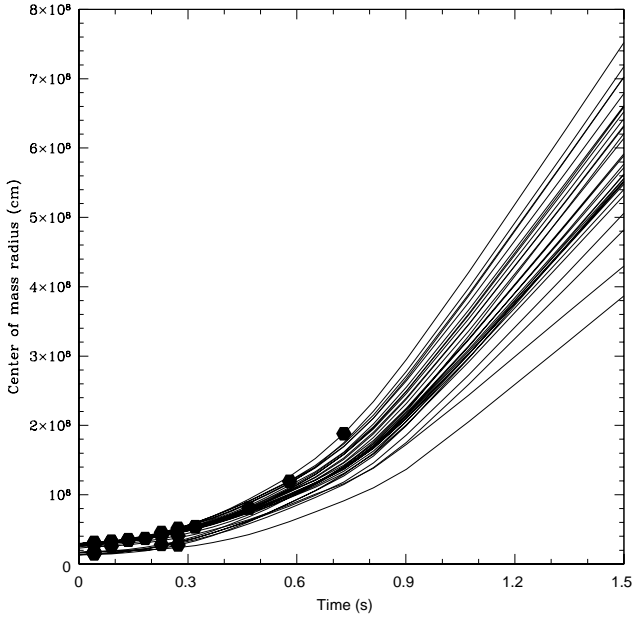}}
\caption{Radii of the centers of mass of the 30 bubbles of model B30U. The
filled symbols over a line mark the time at which interaction with any
other bubble first occurs. }\label{fig6} \end{figure}

\begin{figure}
\resizebox{\hsize}{!}{\includegraphics{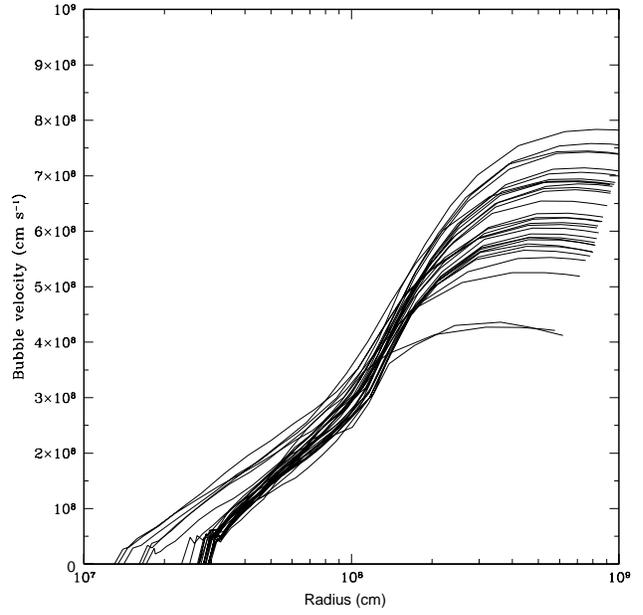}}
\caption{Radial velocity of the bubbles as a function of the radii of their 
centers of mass,
for model B30U.}\label{fig7}
\end{figure}

\subsection{Reference case: 30 bubbles}

\begin{figure}
\resizebox{\hsize}{!}{\includegraphics{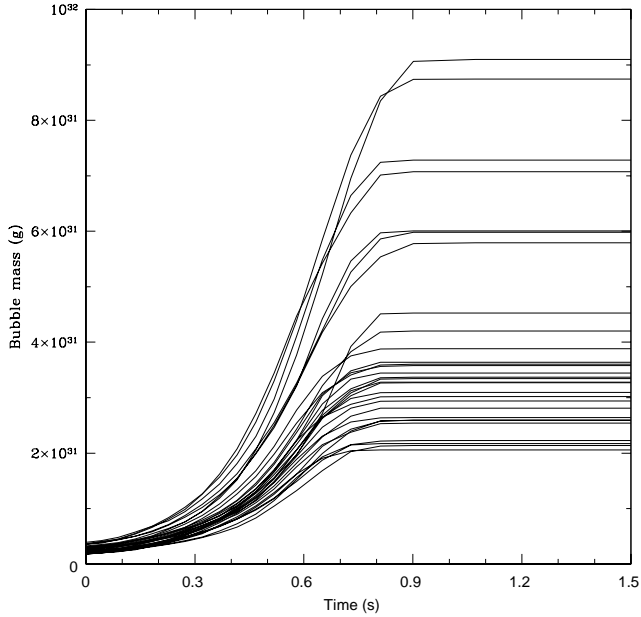}}
\caption{Masses of the bubbles in model B30U. 
}\label{fig8}
\end{figure}

\begin{figure}
\resizebox{\hsize}{!}{\includegraphics{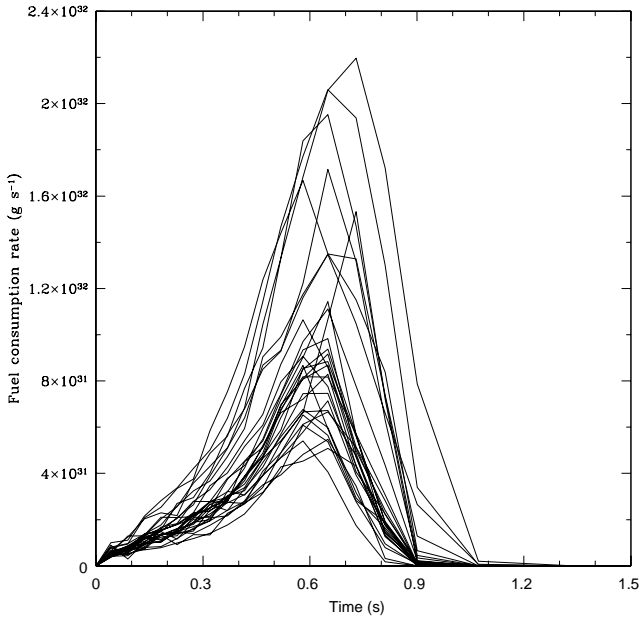}}
\caption{Fuel consumption rate of the bubbles, for model B30U. 
}\label{fig9}
\end{figure}

\begin{figure}
\resizebox{\hsize}{!}{\includegraphics{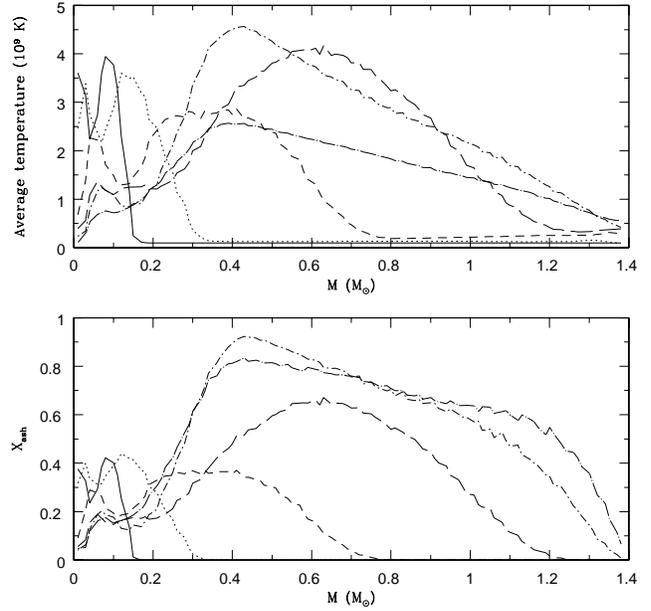}}
\caption{Temporal evolution of temperature (top graph), and burnt mass fraction
(bottom graph), for model B30U. The times shown are: 0 (solid line), 0.2,
(dotted line), 0.4 (short-dashed line), 0.6 (long-dashed line), 0.8 (dot
short-dashed line), and 1.0 s (dot long-dahsed line). The plotted
quantities have been averaged over spherical shells.
}\label{fig10} 
\end{figure}

\begin{figure}
\resizebox{\hsize}{!}{\includegraphics{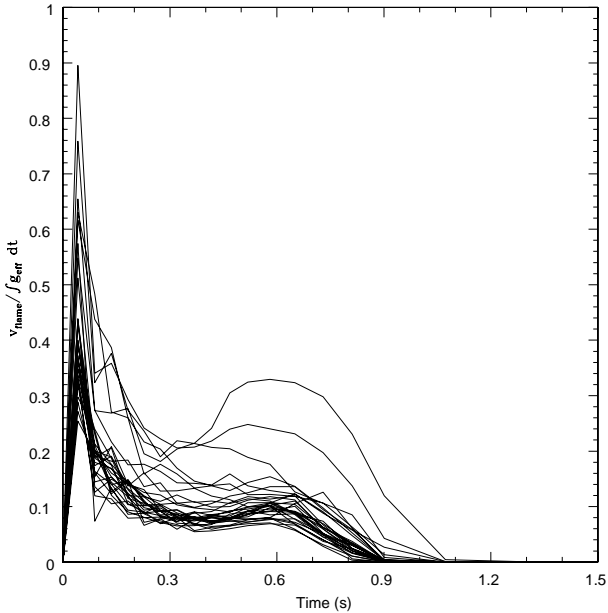}}
\caption{The time evolution of the ratio of the local flame velocity to the 
non-linear Rayleigh-Taylor velocity for each bubble of model B30U. 
}\label{fig11}
\end{figure}

The evolution of the bubbles is governed by the competition between three
processes: 
\begin{itemize}
\item volume increase due to flame propagation and pressure equilibration
with its surroundings,
\item merging of bubbles, and
\item surface deformation due to hydrodynamical instabilities.
\end{itemize}
We start by describing the evolution of model B30U (starting
from 30 bubbles of nearly equal size), whose main trend can be seen in
Figs.~\ref{fig5} to \ref{fig11}.

It takes about 0.3~s for the bubbles to start floating, from then on they follow 
an accelerated motion outwards to reach their terminal velocity before 1~s
(Fig.~\ref{fig6}). The final velocity of the bubbles ranges from 4 to
$7.6\times10^8$~cm~s$^{-1}$(Fig.~\ref{fig7}). Generally speaking, the bubbles that
achieve the largest velocity are those whose initial locations are farther away 
from the center of the white dwarf, although there are exceptions. 
The deformation induced by the interaction among bubbles is a
factor that feeds the hydrodynamic instabilities with a rich spectrum of
scalelengths. This interaction among bubbles begins soon after runaway 
(Figs.~\ref{fig5} and \ref{fig6}).
Rayleigh-Taylor mushrooms are well developed after a few tenths
of a second, increasing the 
net fuel consumption rate, which
peaks between 0.6 and 0.8~s, when the density of the bubbles has decreased
below $10^8$~g~cm$^{-3}$. The local flame velocity in each bubble (defined as
$\left(\D M_\mathrm{b}/\D t\right)/4\pi R_\mathrm{b}^2\rho_\mathrm{b}$)
reaches its maximum at about the same time, and ranges from 400 to 700
km~s$^{-1}$. At the last time
represented in Fig.~\ref{fig5} the merging is so advanced that the
bubbles have lost their individual identity and the nuclear ashes form a
continuum medium, which is the main component in between radii $10^8$ and
$5\times10^8$~cm. The small wavelengths that were present at earlier times
have by then erased, and the overall appearance of the hot region is
dominated by 7-8 large plumes. The central zone of the exploding white dwarf
is occupied by cold unburnt C-O rich matter.

The diversity in the history of the different bubbles
shows up also in the evolution of their growth rates (Fig.~\ref{fig8}). 
A small departure in the size of the bubbles at the beginning is translated
into a large difference in their final mass. The peak in the fuel consumption
rate of each bubble can range over factor of five (Fig.~\ref{fig9}), the
largest values being attained in those bubbles initially located at 
lower altitude. Close to the center, the effective gravity 
is proportional to the radius, which implies that the floatability 
is lower than that of the bubbles born at larger radius.
On the other hand, all the bubbles are already big enough at the beginning of
the hydrodynamical simulation for the Stokes and drag forces not to play an
important role in their evolution. Thus, bubbles born far away from the center
float earlier and grow faster in physical size, but not in mass, with respect
to those born close to the center. The latter, in turn, can sustain 
their large fuel consumption rate during a 
long time, while they remain at high densities. 

The evolution of the averaged (over spherical shells) profiles of temperature 
and ash mass fraction is shown in Fig.~\ref{fig10}. In 
spite of the large departure of the models from 
spherical symmetry, these averages still help to understand
the evolution of the deflagration in our simulations. By looking at the
temperature and composition plots, one can see that during the first 0.4~s the
hot burnt region floats away from the center and spreads over a wide range of radii, while the amount of consumed fuel grows only moderately. At 0.6~s
(long-dashed curve in Fig.~\ref{fig10}) there
is a sudden rise of the temperature and of the ash content, but it occurs  halfway the
total mass of the white dwarf. By 0.8~s (dot short-dashed line in
Fig.~\ref{fig10}) the combustion is also propagating inwards, where the density is
higher. Finally, at the last time shown combustion has almost
stopped and, at the same time, there has been a slight redistribution of the 
ashes towards larger radii.

The onset of the Rayleigh-Taylor instability, and its feedback on the flame
behaviour around the surface of the bubbles can be appreciated from 
Fig.~\ref{fig11}. It shows the ratio between the local flame 
velocity in each bubble and the Rayleigh-Taylor velocity in
the nonlinear regime
\footnote{Note that the gravity that acts on the bubble is a function of time
and thus simple estimates of the Rayleigh-Taylor timescale based on a constant
value of $g$ can be wrong by an order of magnitude (see, e.g., Niemeyer et 
al. \cite{nhw96})}, $v_\mathrm{RT} = \int{g_\mathrm{eff}\D t}$, 
 as a function of time. Three
phases can be outlined from this figure. At early times ($t\la 0.1-0.2$~s) the
flame propagation is not driven by the Rayleigh-Taylor instability, which shows
up as a lack of correlation between the flame velocity and $v_\mathrm{RT}$. From
this time up to $\sim0.6-0.7$~s the velocity of the flame scales  with the
Rayleigh-Taylor velocity, and the ratio $v_\mathrm{flame}/v_\mathrm{RT}$ remains
almost constant with a value within the range 0.07-0.14, depending on the 
bubble.
This second phase is skipped by a few bubbles, those which stay closest to the
center and which can be identified in Fig.~\ref{fig11} by having the largest
ratios. Finally, at still later times the fuel consumption rate starts 
decreasing due to the drop in
density, and the flame velocity decouples again from the developement of the Rayleigh-Taylor
instability.

The final output of the explosion is given in Table~\ref{tab2}, together with 
the rest of computed models.
Model B30U has been followed for a long time after the end of
the combustion phase of the explosion, to gain insight into the evolution of the
ejecta until homology sets in. As can be seen in Table~\ref{tab2}, this model 
was integrated over 19,000 hydrodynamical time steps, until a time 
$t_\mathrm{last} \sim 5,5$~hours. The explosion produced $0.43~M_{\odot}$ of
$^{56}$Ni, and the ejected material acquired a kinetic energy at infinity of
$0.43\times10^{51}$~ergs. Both quantities are within the ranges expected in a
typical Type Ia supernova, although the kinetic energy is at the low end of
the allowed range. The performance of the SPH code can be evaluated through its
capability to conserve integral quantities such as energy, momentum
and angular momentum (last three columns in Table~\ref{tab2}). As can be seen,
the momentum and angular momentum are, for practical purposes, perfectly 
conserved, but the code does not follow as efficiently the transformation of
energy from one kind to another. The total energy is not exactly conserved, the
amount of energy {\sl lost} in the whole simulation being about the 
same as the (low) energy deposited in the bubbles at
$t=0$, in the initial incineration of the particles to NSE (see last column of
Table~\ref{tab1}).

\begin{table*}
\centering
\caption[]{Results of the hydrodynamical simulations.}
\label{tab2}
\begin{tabular}{lccccccc}
\hline
\noalign{\smallskip}
Model & $E^{\mathrm{a}}$ & $M \left(^{56}\mathrm{Ni}\right)$ &
$t_\mathrm{last}$ & $I$ & $\Delta E^{\mathrm{b}}$ &
$\Delta p^{\mathrm{c}}$ & $\Delta L^{\mathrm{d}}$ \\  
& ($10^{51}$~erg) & ($M_\odot$) & 
(s) & (Time steps) & ($10^{51}$~erg) & ($10^9~M_{\odot}\mathrm{cm~s}^{-1}$) & 
($10^{17}~M_{\odot}\mathrm{cm}^2\mathrm{s}^{-1}$) \\ 
\noalign{\smallskip}
\hline
\noalign{\smallskip}
B30U & 0.43 & 0.43 & 19480. & 19000 & $-0.07$ & $4\times10^{-9}$ &
$2\times10^{-6}$ \\
B10U & 0.05 & 0.24 & 1.05 & 1800 & $-0.03$ & $3\times10^{-8}$ & $10^{-9}$ \\
B07U & $-$ & 0.21 & 12.17 & 6100 & $-0.06$ & $5\times10^{-8}$ &
$6\times10^{-9}$ \\
B06U & $-$ & 0.19 & 9.11 & 4400 & $-0.07$ & $3\times10^{-8}$ & 
$5\times10^{-9}$ \\
B90R & 0.45 & 0.44 & 2.20 & 5700 & $-0.07$ & $10^{-8}$ & $4\times10^{-9}$ \\
\noalign{\smallskip}
\hline
\end{tabular}
\begin{list}{}{}
\item[$^{\mathrm{a}}$] Final energy (gravitational+internal+kinetic) of the
ejected mass at the last computed time, $t_\mathrm{last}$.
\item[$^{\mathrm{b}}$] Accumulated error in the total energy at the last 
computed model, after $I$ time steps.
\item[$^{\mathrm{c}}$] Total momentum at the last computed model, normalized
taking as a characteristic velocity of the ejecta $10^9$~cm~s$^{-1}$.
\item[$^{\mathrm{d}}$] Total angular momentum at the last computed model,
normalized taking as a characteristic velocity of the ejecta 
$10^9$~cm~s$^{-1}$, and a characteristic radius of $10^8$~cm.
\end{list}
\end{table*}

\subsection{Dependence on the number and size of the bubbles}

The number of bubbles ignited at $t=0$ turns out to be a very influential
parameter with respect to the evolution and final output of the explosion. From the
results displayed in Table~\ref{tab2}, two trends can be realised: First, for a
small number of bubbles ($N_\mathrm{b} \la 30$) the quantity of $^{56}$Ni
produced and the amount of nuclear energy released grow with $N_\mathrm{b}$.
Indeed, models B06U and B07U remained bound at the end of nuclear burning, and
model B10U became only marginally unbound, but even in this case no more than a
few percent of the mass of the white dwarf would be able to overcome the
gravitational barrier. Nevertheless a word of caution must be given 
when the initial number of incinerated bubbles is low. In that case the use 
of a constant baseline velocity for the subgrid is less justified because the 
regulating mechanism based on flame surface self-adjustement does not work so  
well as in the case of a large number of bubbles displaying a larger surface. 
Thus, the critical number of igniting regions that might result in a 
successful explosion could be slightly different from what was obtained in our 
calculations. This point will be explored through a parameter study and reported in a forthcoming publication (Bravo \& Garc\'\i a-Senz 2005, in preparation). 
Second, for a larger number of bubbles ($N_\mathrm{b} \ga
30$) the explosion converges to a unified solution, and the output is quite
insensitive to both the precise value of $N_\mathrm{b}$ and the size spectrum of
 the bubbles. Thus, models B30U and B90R give rise to almost the same
explosion. This result is remarkable, as it marks the onset of
convergence to an {\sl homogeneous} supernova explosion, independently of the 
precise initial conditions. Recall that homogeneity is a  
 primordial observational property of SNIa. 

\begin{figure*}
\centering
\resizebox{\hsize}{!}{\includegraphics{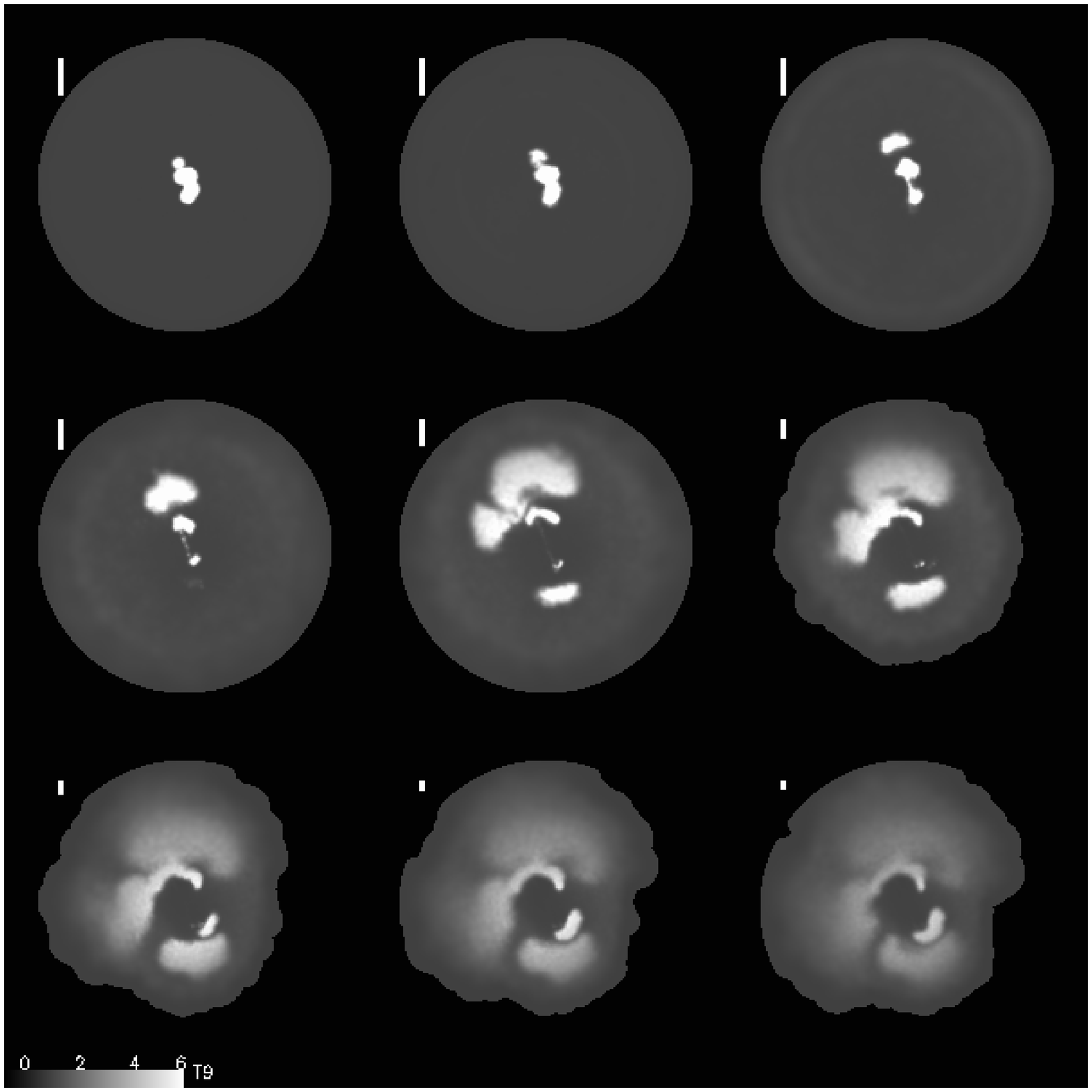}}
\caption{Snapshots of the temperature distribution in a meridian plane of
model B07U at times from 0 to 1.05 s, in steps of $\sim0.13$~s. The 
temperature scale is shown at the bottom left of the image, while
the length scale is shown at the top left of each snapshot (the
length of the vertical bar is equivalent to 200~km).
}\label{fig12} 
\end{figure*}

\begin{figure*}
\centering
\resizebox{\hsize}{!}{\includegraphics{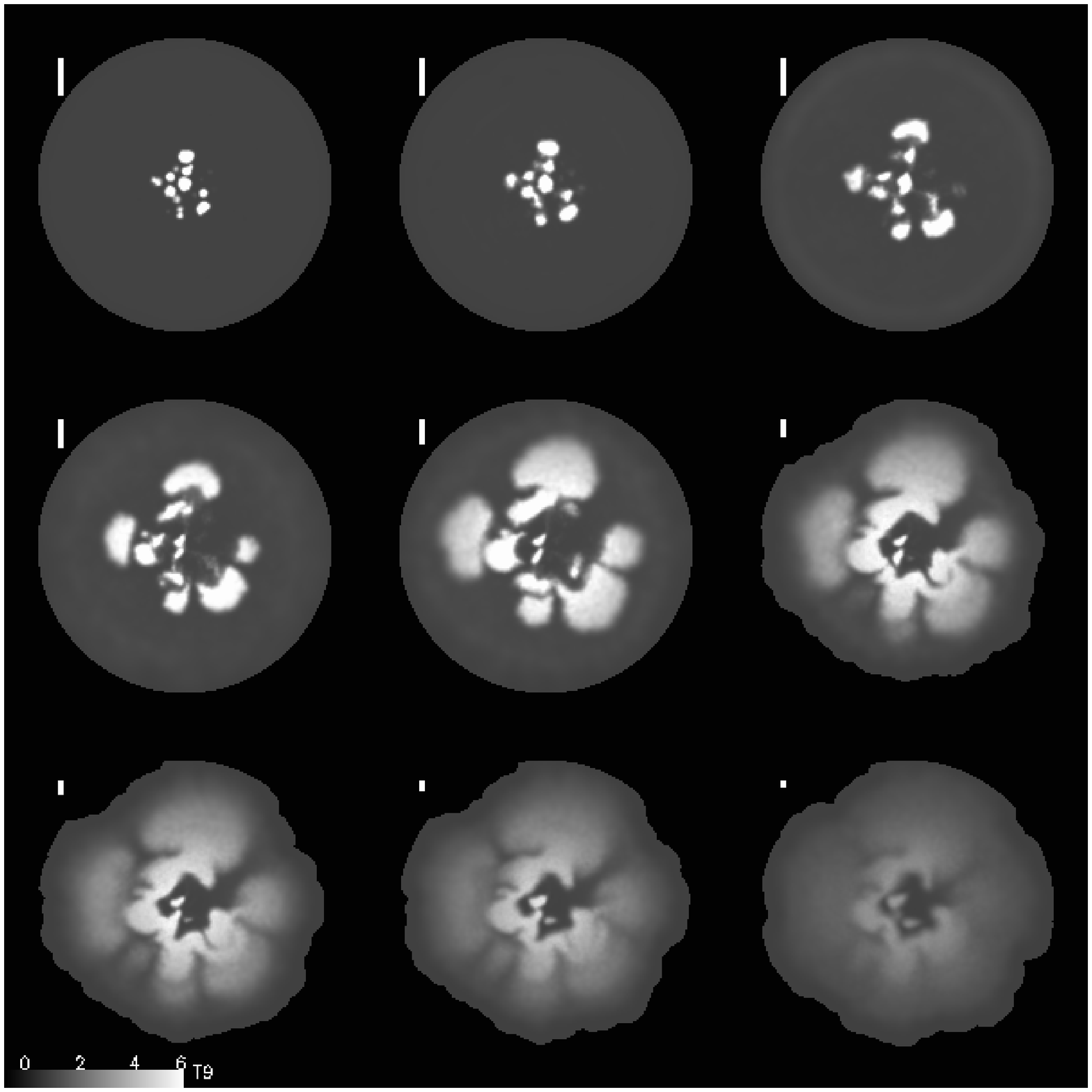}}
\caption{Snapshots of the temperature distribution in a meridian plane of
model B90R at times from 0 to 1.10 s, in steps of $\sim0.14$~s. The 
temperature scale is shown at the bottom left of the image and 
the length scale is shown at the top left of each snapshot (the
length of the vertical bar is equivalent to 200~km).
}\label{fig13} 
\end{figure*}

\begin{figure}
\resizebox{\hsize}{!}{\includegraphics{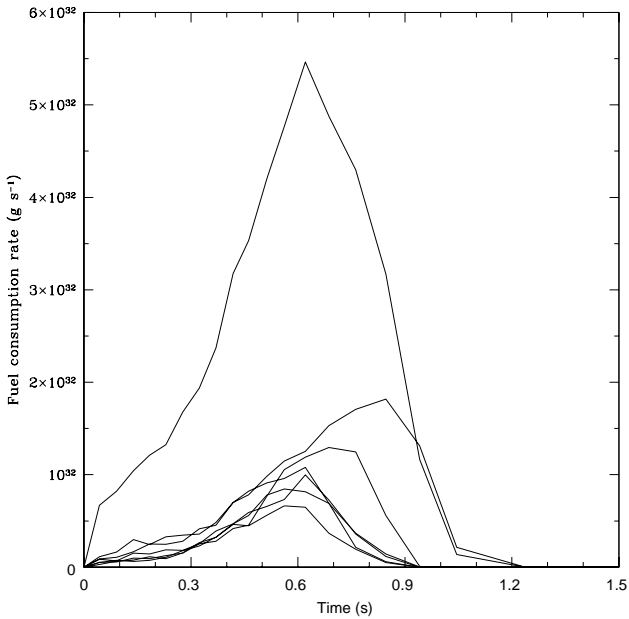}}
\caption{Fuel consumption rate of the bubbles for model B07U. 
}\label{fig14}
\end{figure}

\begin{figure}
\resizebox{\hsize}{!}{\includegraphics{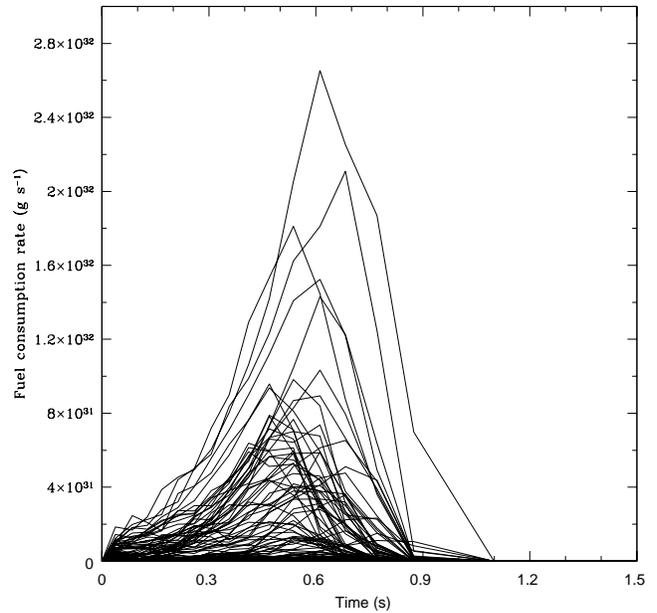}}
\caption{Fuel consumption rate of the bubbles, for model B90R. 
}\label{fig15}
\end{figure}

\begin{figure}
\resizebox{\hsize}{!}{\includegraphics{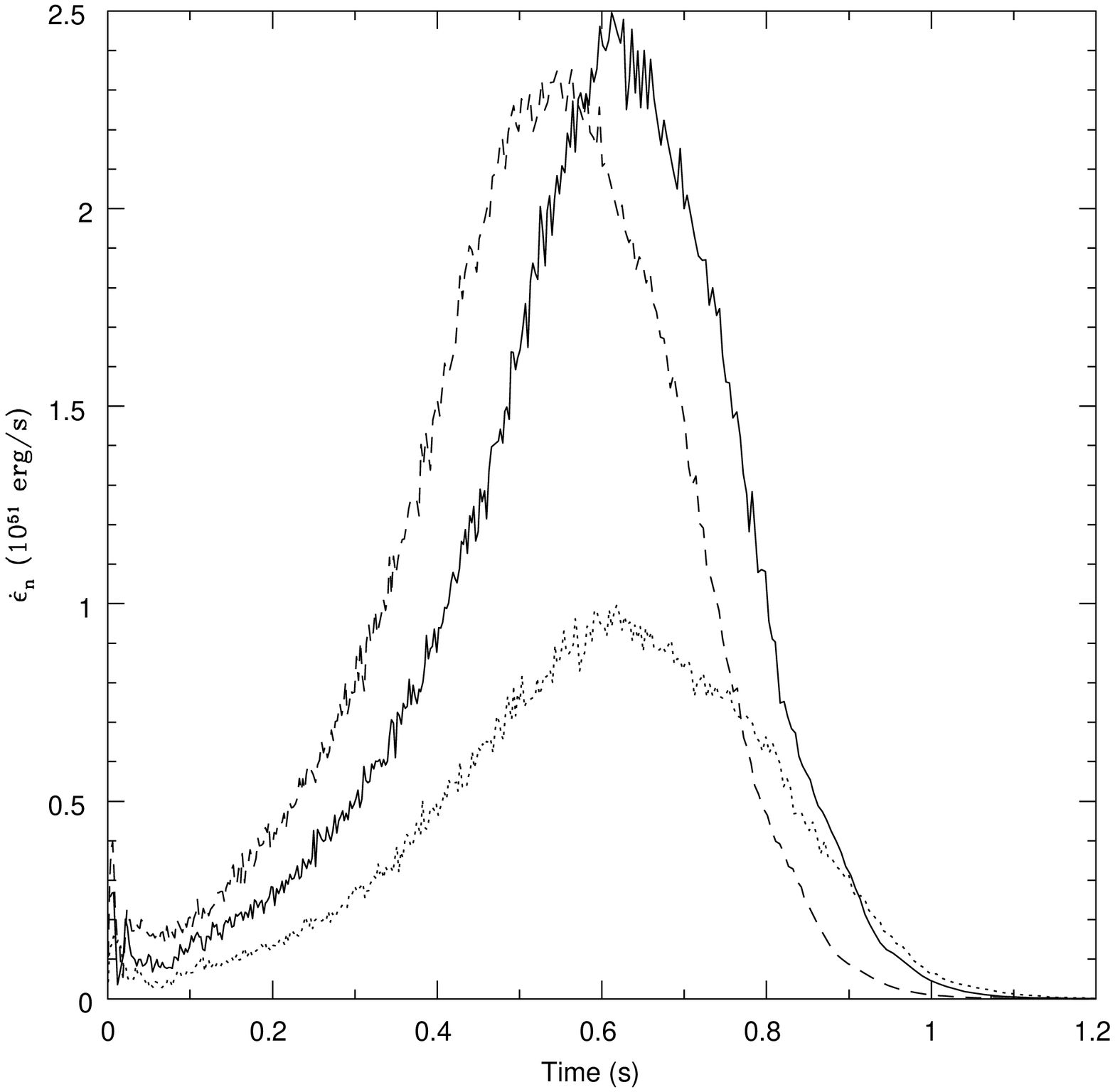}}
\caption{Nuclear energy generation rate as a function of time. Solid line:
model B30U, dashed line: model B90R, dotted line: model B7U.}\label{fig16}
\end{figure}

\begin{figure}
\resizebox{\hsize}{!}{\includegraphics{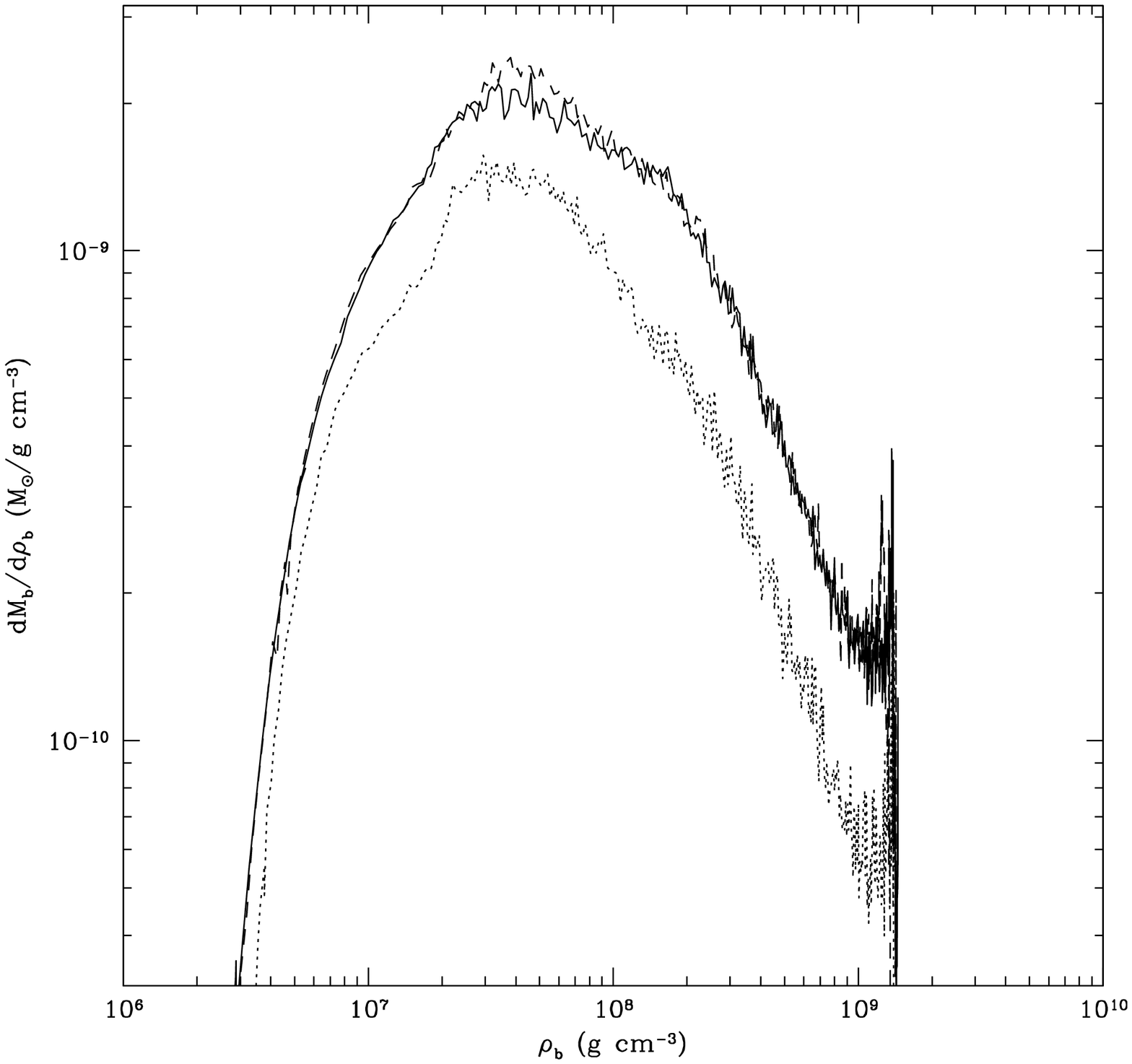}}
\caption{Total fuel consumption rate as a function of density. Solid line:
model B30U, dashed line: model B90R, dotted line: model B7U.
}\label{fig17}
\end{figure}

In the following, we describe the main trends of
models B07U and B90R, as representative of the two groups of models identified
before. The evolution of both models is exemplified in Figs.~\ref{fig12} to
\ref{fig17}. 

It is worth comparing the thermal structure evolution of models B07U
(Fig.~\ref{fig12}) and B90R (Fig.~\ref{fig13}) with that of model B30U
(Fig.~\ref{fig5}). The most evident difference between 
model B07U and the other two is the large departure from spherical symmetry
of the first one. In B07U, there is a dominant bubble (the one located upwards in
the picture), that grows fast, floats the most rapidly, and determines the
overall evolution of the explosion. This can also be deduced from
Fig.~\ref{fig14}, which shows the fuel consumption rate of each bubble. 
The
dominant bubble is subjected to hydrodynamic instabilities, and its 
surface
is deformed sooner. The subsequent increase of flame
surface largely compensates for the expansion effect on nuclear reactions. The
dominant bubble reaches the outer layers of the white dwarf at $\sim0.5-0.6$~s,
when nuclear burning rapidly fades. The bubble then acts like a piston, pushing
against the surrounding material, 
which provokes a break-out and the subsequent outflow of
matter around the surface of the star. 

In model B90R, in contrast, there is not one single bubble that dominates the fuel
consumption rate (Fig.~\ref{fig15}). In this model, not all the bubbles survive
 the first few tenths of a second after initial ignition, because  most of
them are
ingested by the largest blobs. However, the overall evolution is quite similar to
model B30U. The main difference between the evolution of B30U and B90R is that
the latter displays a larger diversity of bubble sizes at any given time. This
implies in turn a different floatability and, thus, a larger range of spatial
scales present in the structure of the 90 bubbles model. Hence, the
Rayleigh-Taylor mushrooms develop earlier, and the fine-scale structure is more
complex at late times. 

The rate of nuclear binding energy released by all the bubbles,
$\dot{\epsilon}_\mathrm{nuc}$, is depicted in
Fig.~\ref{fig16} as a function of time, while the fuel consumption rate
summed over all the bubbles is shown in Fig.~\ref{fig17} as a function of
density. The temporal evolution of $\dot{\epsilon}_\mathrm{nuc}$ is similar for
all the models, although the rate is noticeably smaller for the explosions which start from
a small number of bubbles. On the other hand, the explosions starting from 30 and
90 bubbles differ only in the moment at which the maximum rate is attained; 
this is earlier in B90R by about 20\% in time. Figure~\ref{fig17} shows how most
of the fuel is burnt at moderate densities ($\rho_\mathrm{b}$ in the interval
$2\times10^7 - 2\times10^8$~g~cm$^{-3}$) in all the calculations.

The long-term evolution of the models that fail to explode will be the subject
of a forthcoming paper. In these models, the released energy will result in
expansion followed by recontraction of the white dwarf. However, the chemical
profile
of the infalling matter will be quite different from what is inferred in current
1D
pulsating delayed detonation models. Some hints of the possible outcome of such
a
situation have already been presented in Bravo \& Garc\'\i a-Senz~(\cite{bgs04}).

\subsection{Nucleosynthesis}

The detailed nucleosynthesis produced in the successful
explosions has been computed with a post-processing code. The results, which can 
be found in Table~\ref{tab3}, reveal that models B30U and B90R produce
virtually the same species and in the same quantities. 

\begin{table}
\centering
\caption[]{Nucleosynthesis productivity, in solar masses.}
\label{tab3}
\begin{tabular}{lcc|lcc}
\hline
\noalign{\smallskip}
 & B30U & B90R & & B30U & B90R \\ 
\noalign{\smallskip}
\hline
\noalign{\smallskip}
$^{12}$C  & 0.282 & 0.276 & $^{55}$Mn & 8.4E-4 & 8.0E-4 \\
$^{16}$O  & 0.387 & 0.389 & $^{54}$Fe & 0.044 & 0.041 \\
$^{20}$Ne & 0.027 & 0.027 & $^{56}$Fe & 0.482 & 0.489 \\
$^{22}$Ne & 0.010 & 0.010 & $^{57}$Fe & 9.8E-3 & 0.010 \\
$^{24}$Mg & 0.013 & 0.014 & $^{59}$Co & 2.9E-4 & 3.0E-4 \\
$^{28}$Si & 0.045 & 0.048 & $^{58}$Ni & 0.046 & 0.043 \\
$^{32}$S  & 0.011 & 0.010 & $^{60}$Ni & 0.023 & 0.023 \\
$^{36}$Ar & 1.8E-3 & 1.4E-3 & $^{61}$Ni & 4.8E-4 & 5.0E-4 \\
$^{40}$Ca & 1.8E-3 & 1.3E-3 & $^{62}$Ni & 1.4E-3 & 1.4E-3 \\
$^{44}$Ca & 2.6E-5 & 2.6E-5 & $^{65}$Cu & 3.5E-6 & 3.6E-6 \\
$^{48}$Ti & 4.6E-5 & 4.3E-5 & $^{64}$Zn & 4.3E-4 & 4.5E-4 \\
$^{52}$Cr & 2.7E-4 & 2.2E-4 & $^{66}$Zn & 3.7E-5 & 3.8E-5 \\
\noalign{\smallskip}
\hline
\end{tabular}
\end{table}

\begin{figure}
\resizebox{\hsize}{!}{\includegraphics{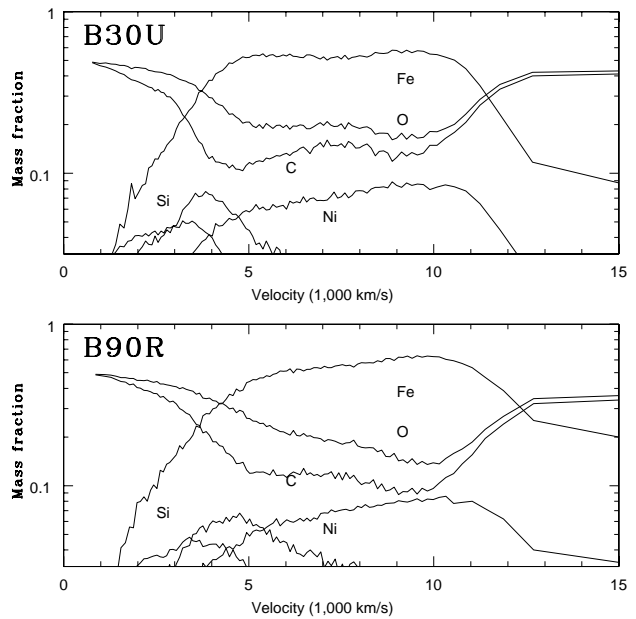}}
\caption{Final distribution of elements in velocity space. Top: model B30U,
bottom: model B90R. The composition corresponds to the final products after
radioactive disintegrations.
}\label{fig18} 
\end{figure}

The most salient nucleosynthetic feature is the large amount of unburnt C and
O. As shown in Fig.~\ref{fig18}, this C-O rich matter is present at any distance
from the center of the ejecta, although it concentrates preferentially in the
innermost and in the outermost
layers. The presence of carbon and oxygen in the central layers of 3D
deflagration models has sometimes been regarded as a failure of the pure 
deflagration scenario (Gamezo et al.~\cite{gkochr03}), but it is not clear at 
all whether or not they would be actually detectable in current observations of 
SNIa (Baron et al.~\cite{blh03}). 

The isotopic composition of Fe-peak nuclei presents moderate excesses of
$^{54}$Fe, $^{58}$Ni and $^{60}$Ni. Our post-processing calculations start from
an initial composition of 49\% $^{12}$C, 49\% $^{16}$O, and 2\% $^{22}$Ne.
Further neutronization is provided by electron captures on NSE matter at high
density. Basically, this affects the matter burned in the bubbles at the
beginning of the hydrodynamical simulation (i.e. at $t=0$). The time scale for
the decrease of the density of the bubbles is $\sim 0.3$~s, but in that time the mass of
the bubbles hardly increases by a factor $\sim2$ above its initial value
(Fig.~\ref{fig8}). The final electron molar number of the particles incinerated
at $t=0$ reaches a value of $\sim0.468$~mol~g$^{-1}$, and the composition
corresponding to freezing-out from NSE at this $Y_\mathrm{e}$ is
dominated by the same neutronized nuclei as mentioned above, plus $^{56}$Fe. 

The amount of intermediate-mass elements (IME) ejected in the explosions is 
rather small (Table~\ref{tab3}), and their distribution in velocity space is
excessively smeared and too close to the center (Fig.~\ref{fig18}) for a Type
Ia supernova. Indeed, the presence of $^{56}$Ni ($^{56}$Fe after the 
radioactive decays) in the external layers
would produce quite distinctive spectral features near maximum, which 
remain undetected in
SNIa so far. The scarcity in the production of IME could be in part related to
the rough description of flame propagation at low densities
($\la10^7$~g~cm$^{-3}$), which is a weakness common to most 3D SNIa hydrocodes. 

The worst flaw of the present models is the high degree of mixing of the
elements in velocity space. The stratification of the ejecta is a feature
strongly suggested by the observations, only allowing  a moderate amount of
mixing. The absence of such stratification in
the models is a direct consequence of the formation of large plumes of ashes due
to hydrodynamic instabilities. As strong radial mixing is a common feature of  
all 3D deflagration models
computed by the astrophysical community (e.g. G03 and RHN), it casts  
serious doubts on the soundness of pure deflagrations as 
models of Type Ia supernovae.

\begin{figure}
\resizebox{\hsize}{!}{\includegraphics{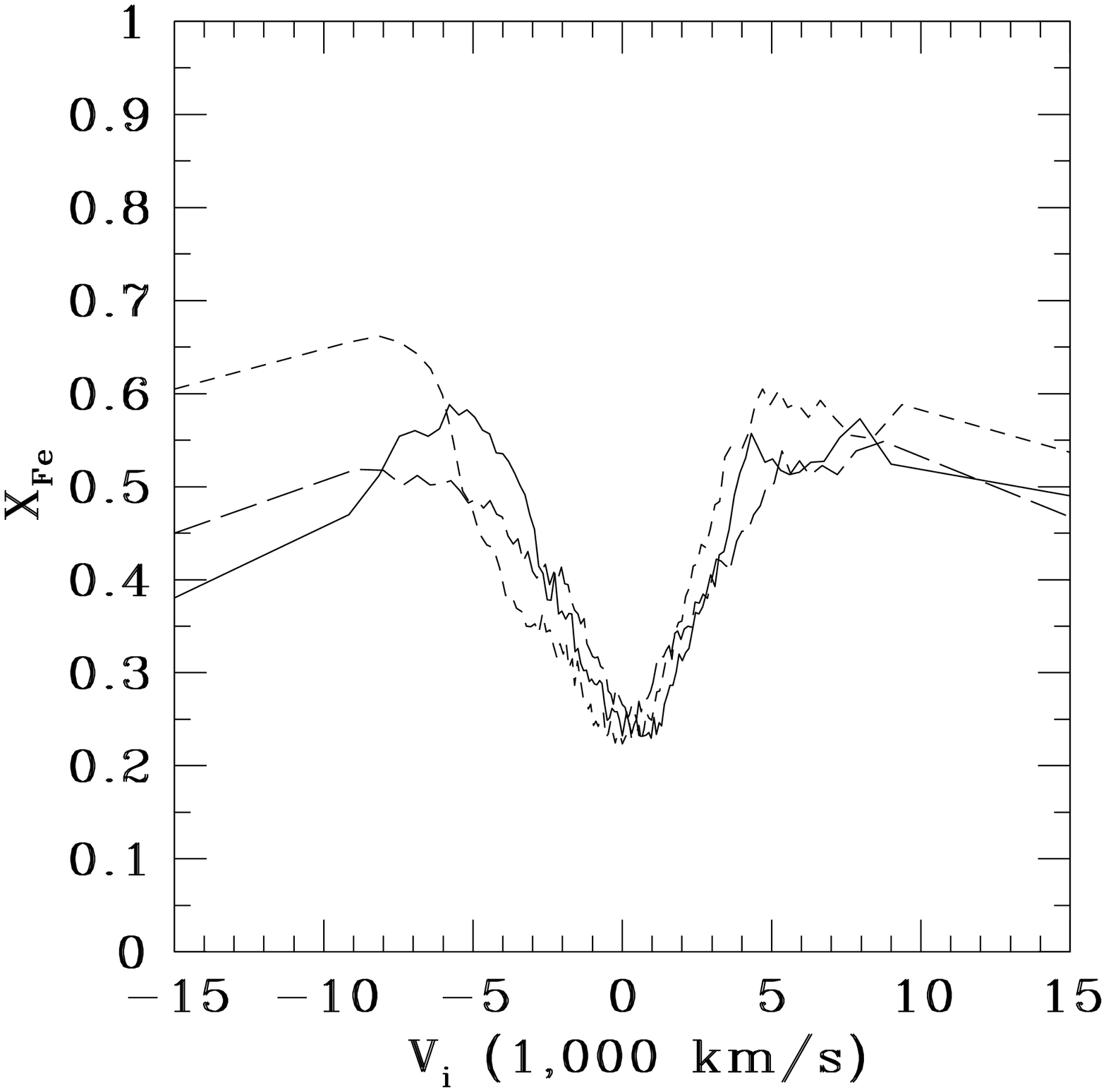}}
\caption{Distribution of iron as a function of the components of the velocity,
model B30U. 
}\label{fig19} 
\end{figure}

\begin{figure*}
\centering
\resizebox{\hsize}{!}{\includegraphics{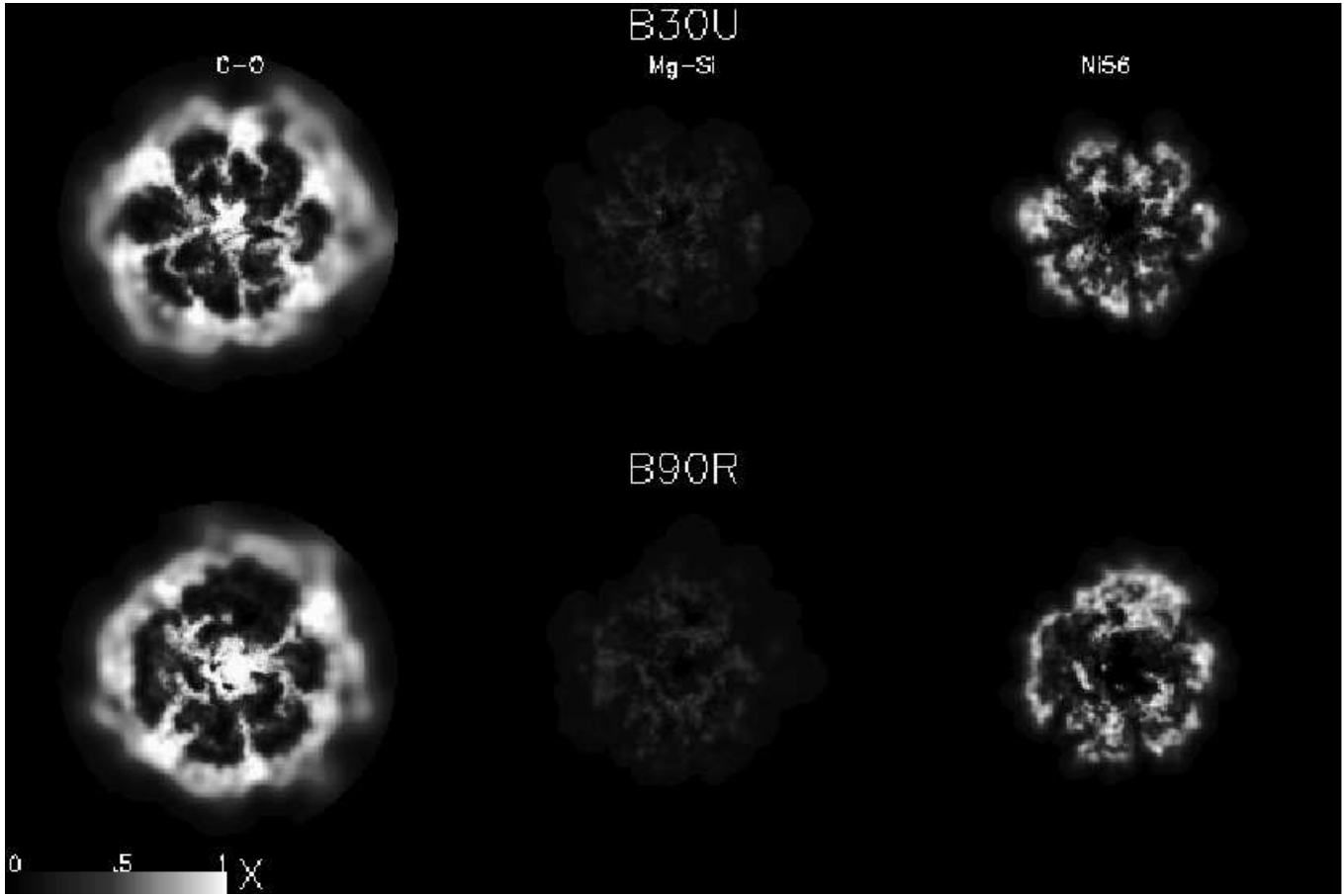}}
\caption{Final (at $t\sim3$~s) chemical composition mapped onto a meridional 
plane, for models B30U (top row) and B90R (bottom row).
}\label{fig20} 
\end{figure*}

As it is the result of a 3D simulation, one should not expect that the material is
ejected with spherical symmetry. However, the distribution of Fe with
respect to different directions of motion is quite symmetric (Fig.~\ref{fig19}),
showing a small departure from the average curve only in the high-velocity tail
of one of the components of the velocity. Such a small
asymmetry would hardly have any observable consequences. On the other hand, the
distribution of elements in physical space is neither homogeneous nor symmetric
(Fig.~\ref{fig20}), especially that of $^{56}$Ni which concentrates in a few 
large pockets. The possible relevance of this heterogeneous distribution of
$^{56}$Ni will be the subject of discussion in the next section.

\section{The long term evolution}

The large computational cost of 3D hydrodynamical calculations is a factor that
severely limits the time range explored in the simulations. Usually, 3D
simulations of SNIa are followed for no more than 1-2~s, and the final model is
expected to be representative of the final output of the thermonuclear 
explosion. Even though by that time the nuclear reactions have quenched and the
chemical composition is already fixed, the density is still high and the dynamical
interactions between different regions in the ejecta are able to change
substantially the distribution of specific kinetic energy. With the aim to
explore these effects, we have followed one of the succesful explosions (B30U)
until an elapsed time of 19480~s (see Table~\ref{tab2}). Our intention is to
try to answer three fundamental questions in this section (within the limits 
imposed by the limited time we have been able to calculate): 
\begin{itemize}
\item When do the ejecta achieve the homologous expansion phase?
\item How can models be extrapolated from $\sim1$~s up to times spanning 
several hours?
\item What other effects become relevant for determining the ejecta structure at
later times?
\end{itemize}

\begin{figure}
\resizebox{\hsize}{!}{\includegraphics{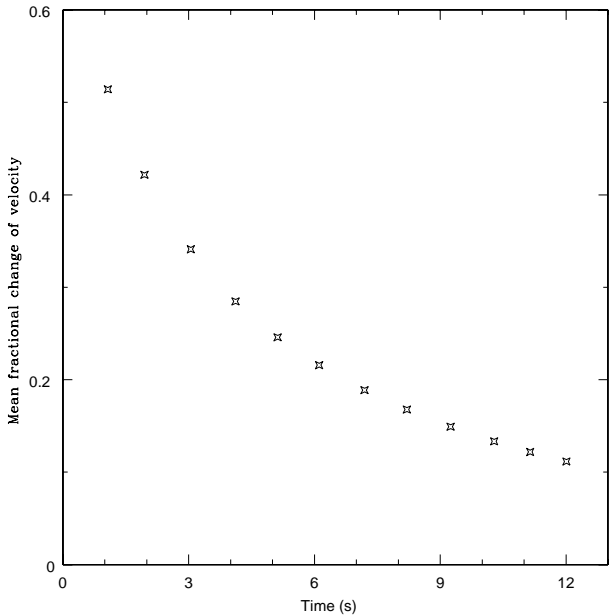}}
\caption{Mean of the fractional change in particles velocity with respect to its
final velocity. A 5\% difference is attained at $\sim16$~s, and a 1\% difference
at $\sim27$~s.  
}\label{fig21} 
\end{figure}

Figure~\ref{fig21} is intended to provide a quantitative answer to the first 
question. We
have computed, for model B30U, the difference between the velocity of each
particle in the last computed model and its velocity at 
different, earlier, times. The
quantity depicted in the figure is the average of these differences averaged over
all the particles. It can be seen that at $t=1$~s the velocities still have
to change, in average, by almost 50\% of their final value. This fractional
change drops to $\sim35$\% at $t=3$~s, and to 10\% at $t=12$~s,  99\%
 of the final velocity, on average, is reached at $t=27$~s. These
figures can be considered representative of any 3D SNIa model, although the precise values can
change with the kinetic energy of the ejecta or with the mechanism of the
explosion. Generally speaking, more energetic explosions evolve faster and
arrive earlier at the homologous expansion phase. For instance, for an explosion
twice as energetic as B30U, the mean velocity would scale as $2^{1/2}$, while
the hydrodynamical time scale would change as $\rho^{-1/2} \propto R^{3/2}
\propto v^{3/2}$, i.e. it would evolve $\sim 1.7$ times faster than B30U. Thus, 
the onset of the
homologous expansion phase is expected to occur at times $\ga 16 - 30$~s.

\begin{figure}
\resizebox{\hsize}{!}{\includegraphics{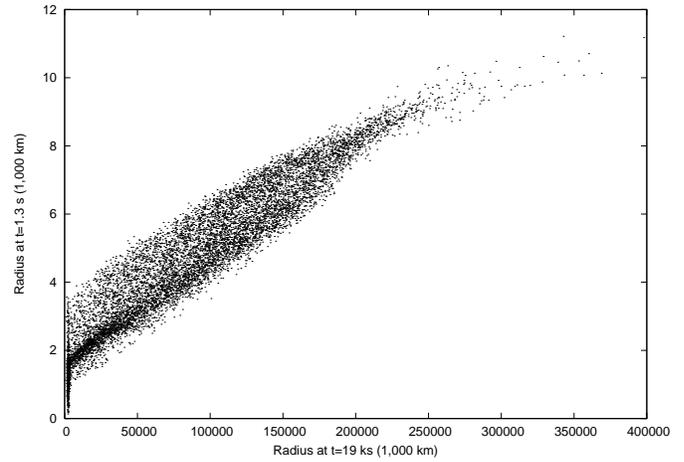}}
\caption{Radius of each SPH particle at $t\sim1$~s vs final radius, for model B30U. 
}\label{fig22} 
\end{figure}

\begin{figure}
\resizebox{\hsize}{!}{\includegraphics{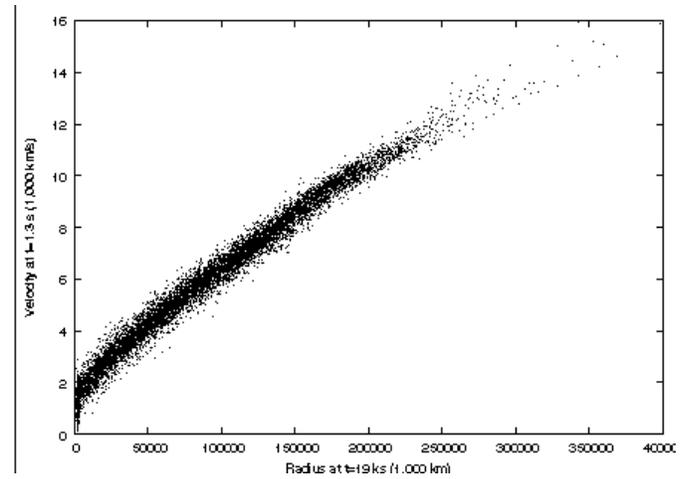}}
\caption{Velocity of each SPH particle at $t\sim1$~s vs final radius, for model B30U. 
}\label{fig23} 
\end{figure}

Very often, 3D simulations cannot be extended much beyond $1-2$~s. In these
cases, even though a detailed nucleosynthetic calculation can be performed, the
final distribution of elements in velocity space remains uncertain. In
principle, one has two ways to extrapolate the known distribution at early times
in order to guess what the final chemical profile will be: either sorting the
chemistry by radius, or by velocity (at $t\sim1-2$~s, the homology
relationship $v\propto r$ does not apply, as we have just seen). 
Figures~\ref{fig22} and \ref{fig23} show the result of both kinds of
extrapolation procedures compared to the actual distribution in the last
computed model of B30U. The relevant quality of the distribution of points shown
in those figures is its
width. Wide distributions (like that in Fig.~\ref{fig22}) are indicative of
important changes in the ordering of the particles, while narrow ones point to
a larger degree of conservation of the order. The conclusion is clear:
sorting by velocity gives results much closer to the final situation than
sorting by radius. 

\begin{figure}
\resizebox{\hsize}{!}{\includegraphics{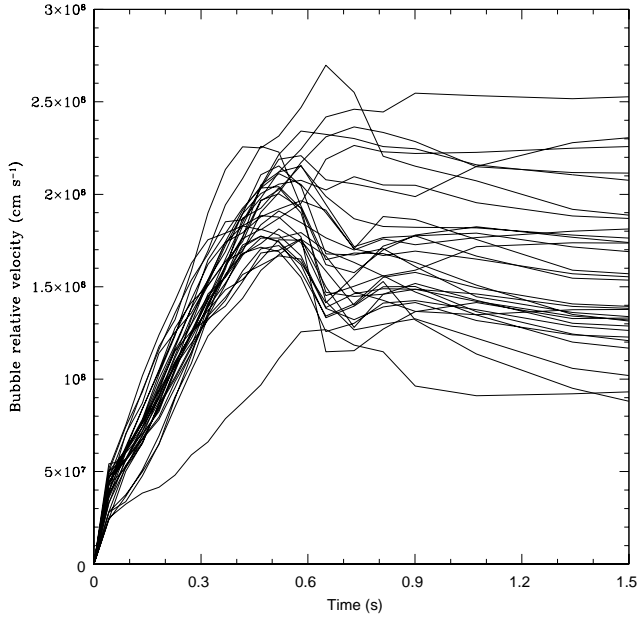}}
\caption{Radial velocity of the bubbles relative to their environment, for model
B30U. 
}\label{fig24} 
\end{figure}

A hint of the reason why sorting by velocities gives better results can be
sketched from inspection of Fig.~\ref{fig24}. There, we have plotted the 
differential velocity of the bubbles with respect to their immediate 
surroundings as a function of time. Bubbles experience a strong acceleration 
due
to buoyancy during the first $0.3-0.5$~s. During this period, the transfer of
momentum to the surrounding material is small, hence the differential velocity of 
the bubbles rises steadily. During the next $\sim0.2$~s, the outermost matter is
accelerated by the bubbles, mainly due to the large increase in their lateral 
size, which makes radial transfer of momentum more efficient. Then, 
a short period of relative deceleration ensues, followed by a stabilization 
of the differential velocity.
In this last phase the bubbles still 
move through the blobs of unburnt C-O, reaching farther regions and slowly
transferring a fraction of their momentum to them.

As we have shown before, the NSE matter expelled in the explosion is mostly 
concentrated in several discrete pockets. The presence of Fe-rich knots in the
outer regions of supernova remnants has already been confirmed in several cases,
particularly in the remnant of the historical Type Ia SN1572. But
the observed knots are much smaller than those found in our 
simulations (Fig.~\ref{fig20}). Indeed, the presence of chemical inhomogeneities
(clumps) in the photosphere of SNIa has been limited by the spectral analysis of Thomas 
et al.~(\cite{tkbb02}). The basic argument is that the presence of huge clumps 
would modify the profile of the SiII absorption line depending on the line of
sight, in contrast with the high degree of spectral homogeneity of typical SNIa.
To remain compatible with the homogeneity of SNIa, Thomas et al. estimated that
the area of the clumps should represent less than 10\% of the area of the
photosphere at the moment of maximum luminosity.

\begin{figure*}
\centering
\resizebox{\hsize}{!}{\includegraphics{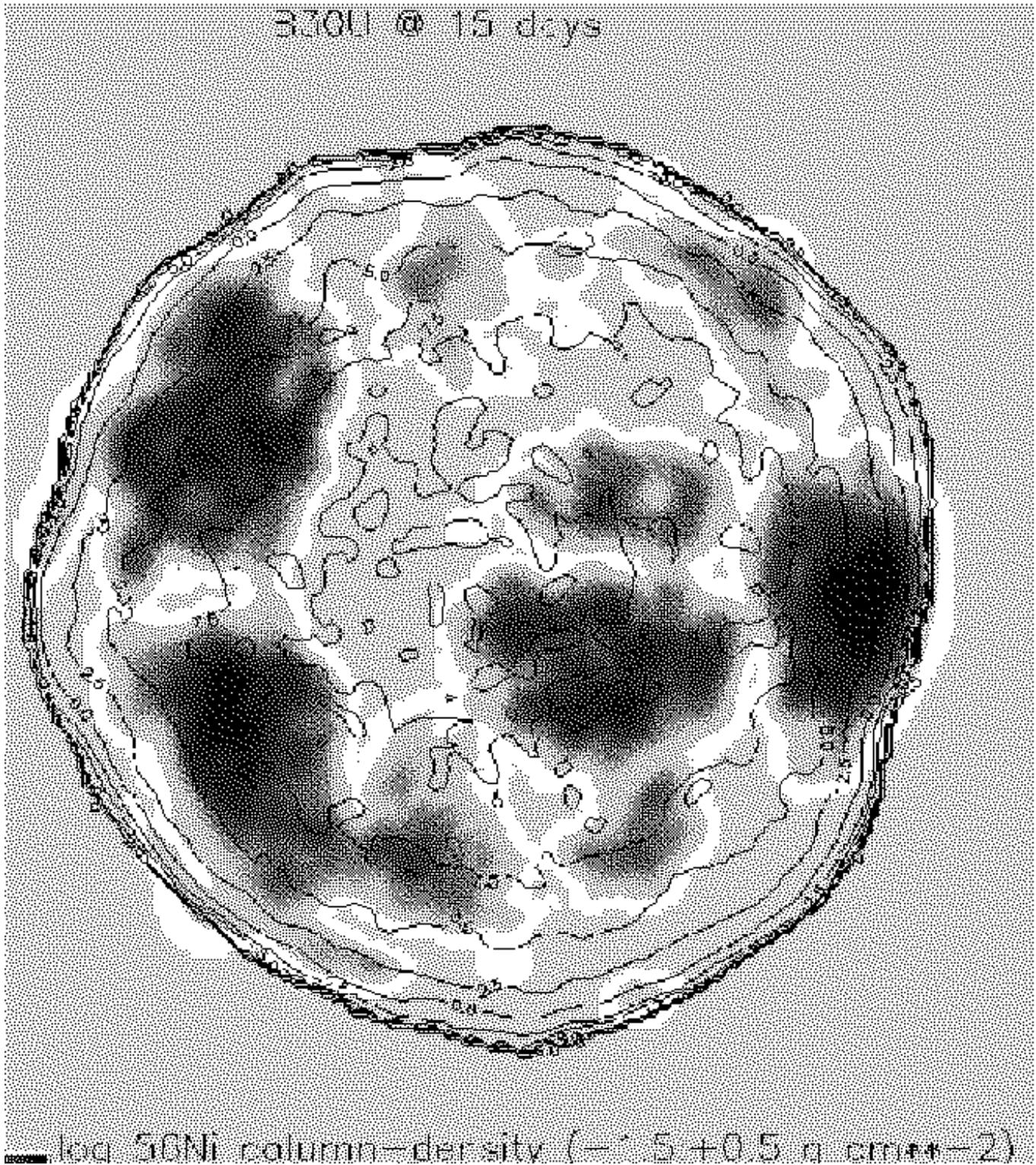}}
\caption{Column density of $^{56}$Ni above the photosphere and contours of 
constant normal velocity at the photosphere, both for model B30U 15 days after 
the explosion. 
}\label{fig25} 
\end{figure*}

Even though we were not able to extend our hydrodynamical simulations 
until such late times 
($\sim15$~days),  we have made a rough estimation of the clumpiness of the ejecta in the following 
way. We have assumed homologous expansion from the last computed model 
($\sim4.5$~hours) up to 15 days. Then we have calculated an approximate location of 
the photosphere, in a given observational direction, by assuming a constant opacity, 
$\kappa\sim0.2$~cm$^2$g$^{-1}$, and computed the column density of material from the
surface inwards, until a prescribed optical depth ($\tau = 2/3$) was reached. The 
results are displayed in Fig.~\ref{fig25}, both for the column density of $^{56}$Ni 
above the photosphere, and for the constant photospheric velocity curves
(note that a $^{56}$Ni column density of 3.3 means that all the
material above the photosphere in that location and in the direction normal to the 
plane of the figure is pure $^{56}$Ni). The velocity curves show a large degree of spherical 
symmetry, with a large velocity gradient towards the limb, due to the 
lower component of the velocity in the observer's direction. The overall appearance
is dominated by 4-5 large clumps of $^{56}$Ni. We believe that this kind of
configuration of $^{56}$Ni clumps is common to all 3D pure deflagation models and,
thus, makes this class of explosion mechanism difficult to reconcile with SNIa 
observations. 
A delayed detonation (DDT) could alleviate this problem by breaking the clumps when the
detonation hits them. The simulation of delayed detonations in 3D is necessary to
ascertain the ability of DDT models to improve the results obtained within the 
pure deflagration scenario.

Another effect we have not taken into account is 
the deposition of the radioactive energy into the ejecta, the so-called
Ni-bubble effect. Depending on the size of the $^{56}$Ni-rich bubbles, the gamma 
photons will be able or not to escape from the bubbles and deposit their energy
either in their own bubble or in the surrounding matter. Thus, the Ni-bubble effect
could make  the problem of the clumps even worse by inducing additional 
expansion of the Ni-rich regions. 

\section{Conclusions}

We have carried out hydrodynamical simulations and calculated the resulting 
nucleosynthesis of thermonuclear 
supernovae within the deflagration paradigm, starting from a few incinerated 
bubbles scattered through the central region of a white dwarf. This ignition 
scenario has recently received  much attention from the supernova community, because
it represents a more natural way of birth of the flame than the previously 
assumed ignition in a central volume. In this scenario, the evolution of the white
dwarf has to be followed in 3D to avoid artificial solutions found previously in 2D 
calculations, as a consequence of symmetry impositions. The most recent 3D 
simulations to which our work can be compared are those from G03 and from the
Munich group (e.g., RHN). In spite of the important differences in the computational
procedure (SPH code in this work, PPM approach in G03 and in RHN), in the description
of the flame (reaction-diffusion equation in our code and in G03, flame tracking 
through advection of a scalar quantity in RHN), in the maximum numerical resolution
(20~km here, 2.6~km in G03, 3.33~km in RHN), in the precise initial conditions,
and in the physics modules, the results of the different groups bear a close
resemblance. The convergence of the gross features of the explosion picture
gives us confidence that our simulations caught the fundamental
properties of deflagrations in white dwarfs. The modest resolution attained in
our calculations has allowed us to undertake a number of simulations that
enlarge considerably the initial conditions explored so far. Moreover, in one 
case we have been able to follow the evolution of the explosion up to several
hours after the thermonuclear burning stops. 

The models starting from $\ga30$ bubbles produced a healthy explosion of quite
uniform characteristics, with nice
amounts of $^{56}$Ni and kinetic energy. However, the deformation of the flame
surface during the first second of the explosion, owing  to the inherent 
hydrodynamical instabilities, gave rise to  important mixing of chemical 
elements
through the ejecta. The absence of chemical stratification appears a serious
problem of current 3D deflagration models of SNIa. Moreover, there are other
drawbacks: an excessive mass of unburnt C-O, a large fraction of which moves at
low velocity, 
close to the center of the ejecta, and the formation of
large clumps of $^{56}$Ni that show up at the photosphere near maximum
brightness, in contrast with what the observational homogeneity of the SNIa
sample suggests (Thomas et al.~\cite{tkbb02}).

Nevertheless, the above models, characterized by a large number of bubbles ignited
nearly-simultaneously, are not the only way a white dwarf can reach thermal
runaway. Leaving aside the central ignition models, we have explored the 
possible initial configurations of igniting bubbles through a  
statistical analytical model.
We have identified two extreme possibilities: a) a set of bubbles with almost 
equal size, and b) a set of bubbles of quite different dimensions whose number 
distribution is nearly independent of
their size. The actual configuration that the white dwarf will have at runaway
will depend on the degree of thermal homogeneity of the central region,
$r\la400$~km. If the thermal gradient around each hot spot is small enough, 
the configuration consisting of bubbles of equal size is preferred; 
otherwise, if the temperature changes by more than
1\% in less than $\sim100$~m, the second kind of configuration would be 
realised.

However, with respect to the results of the explosion, the only relevant
parameter is the number of bubbles ignited at $t=0$. If this number is large
enough ($\ga30$ bubbles in the whole star or $3-4$ per octant), the explosion
converges to the unified solution mentioned before. In contrast, if the number of
igniting bubbles is scarce ($\la2$ per octant) the explosion fails, and an oscillation of the
white dwarf ensues. The final outcome of these 
 oscillations, (if, for instance, they lead to a delayed 
detonation that could still produce a SNIa explosion) will be the
subject of a forthcoming publication.

\begin{acknowledgements}

This research has been partially supported by the CIRIT and MCyT programs
AYA2000-1785, AYA2001-2360, and AYA2002-04094-C03.    

\end{acknowledgements}

\appendix
\section{Statistical approach to the initial distribution of igniting bubbles}

The purpose of this appendix is to develop a statistical study of the size spectrum of
the igniting bubbles located in the central region of a massive white dwarf at
thermal runaway. Specifically, we try to determine the distribution
function of the size, $R_{\mathrm b}$, of the igniting bubbles defined as the 
number of bubbles per unit size interval: $\phi \left(R_{\mathrm b}\right)\equiv 
\D N/\D R_{\mathrm b}$. This distribution function is itself a function of time. 
Our statistical approach is based on the following assumptions:

\begin{enumerate}\renewcommand{\labelenumi}{\bf{\Alph{enumi})}}
\item As a result of convection during the hydrostatic phase there appear  
several hot spots, which we assume to be spherically symmetric and initially 
chemically homogeneous. Instead of treating a hot spot as an isothermal 
sphere, we allow the temperature to be a function of distance to the center 
of the bubble, $R$. Each hot spot is then characterized by its central (peak) 
temperature, $T_0$, and its thermal profile. 
\item Each hot spot evolves adiabatically in place, driven by its own nuclear 
energy generation rate, in a time given by its ignition timescale, $\tau_{\rm
i}$. During this 
time, the temperature profile of the hot spot is modified due to the energy released 
by nuclear reactions, and to thermal conduction within the hot spot (see
Fig.~\ref{fig1}). For simplicity, we consider here that there is no 
interaction between different hot spots. 
\item The statistical distribution of the {\sl initial} peak temperatures of all 
the hot spots can be described by a continuous 
function, $\theta\left(T_0\right)$. The meaning of $\theta$ is the following: 
the 
number of hot spots whose peak temperature is inside the interval $\left[T_0:T_0 
+ \D T_0\right]$ is given by $\D N = \theta\left(T_0\right) \D T_0$. According 
to the above definitions, the desired distribution function of bubble radii is 
given by: $\phi \left(R_{\mathrm b}\right) = \theta\left(T_0\right) 
\left(\D R_{\mathrm b}/\D T_0\right)^{-1}$. The dependence of the radius of the
incinerated 
bubble at any time, $R_\mathrm{b} (t)$, on the {\sl initial} peak temperature of
the hot spot, $T_0$, is the main ingredient of our statistical approach.
\end{enumerate}

We will work with two alternative distribution 
laws of the {\sl initial} peak temperature of the hot spots, 
$\theta\left(T_0\right)$, either a gaussian or a linear law, both of which are 
defined in the interval from $T_{0}^\mathrm{min}$ to $T_{0}^\mathrm{max}$.    
Both distributions are decreasing functions of $T_0$ such that 
$\theta\left(T_{0}^\mathrm{max}\right)=0$. The gaussian distribution function  
is defined as:

\begin{equation}
\theta_{\mathrm G}\left(T_0\right) = 
2{{T_{0}^\mathrm{max}-T_0}\over{B'^2}}{{\exp\left[-\left({{T_{0}^{\mathrm      
max}-T_0}\over{B'}}\right)^2\right]}\over{{1-\exp\left[-\left({{T_{0}^{\mathrm  
max}-T_{0}^\mathrm{min}}\over{B'}}\right)^2\right]}}};
\end{equation}

\noindent 
where $B'=10^8$~K is an arbitrary parameter.
The linear distribution function is defined as:

\begin{equation}
\theta_{\rm L}\left(T_{0,8}\right) = \frac{T_{0}^\mathrm{max} - 
T_0}{2\times10^8}\; ,
\end{equation}

\noindent 
where $T_{0,8}$ is $T_0$ in units of $10^8$K. In what 
follows, we use the values $T_{0}^\mathrm{min}=5\times10^8$~K (the ignition 
timescale below this temperature becomes longer than 4h), and $T_{\rm 
0}^\mathrm{max}=7\times10^8$~K. Thus, once a particular $\theta$ has been chosen  we only need to 
find the
function $\left(\D R_{\mathrm b}/\D T_0\right)$ to be able to determine $\phi$.

At densities $\rho\sim2-3\times10^9$~g~cm$^{-3}$ the ignition timescale can 
be well approximated by 
$\tau_{\mathrm i}\left(T\right)\sim A \left(T/T_{\mathrm b}\right)^{-B}$, 
with $A = 0.0193 s$, $B = 17.97$, and $T_{\mathrm b} = 10^9~{\mathrm K}$ (cf.
see Khokhlov~\cite{k90}, where it is called the induction time). For 
practical purposes, matter incinerates nearly instantaneously whenever its 
temperature exceeds  $T_{\mathrm b}$. Central ignition of a hot spot 
occurs at a time  
$t = \tau_{\rm i}\left(T_0\right) - \tau_{\rm i}\left(T_{\mathrm b}\right)$ 
and thereafter the bubble will grow in mass and size. The first ignition of
any hot spot is attained at a time 
$t = \tau_{\rm i}\left(T_{\rm 
0}^\mathrm{max}\right) - \tau_{\rm i}\left(T_{\mathrm b}\right)$.
As can be 
seen in Fig~\ref{fig2}, the thermal profile of each individual bubble at central
ignition (i.e. when the central temperature becomes equal to $T_{\mathrm b}$)   
can be fitted by a decreasing function of 
$R$: 

\begin{equation}
T(R) = T_{\mathrm b}\exp\left[-\left(R/R_0\right)^2\right]\; , 
\end{equation}

\noindent
where $R_0$  
is a characteristic lengthscale\footnote{Even though the results shown in
Fig.~\ref{fig2} were computed using an initial thermal profile given by
Eq.~1, we have repeated the calculations with different profile shapes and
we always obtained the same qualitative results}.

Given the short timescales involved, there are only two mechanisms that can 
contribute to the growth of a bubble: conductive flame 
propagation and spontaneous flame propagation. One can expect that close to the 
center of the blob (large temperature, shallow gradients)  spontaneous flame 
propagation will dominate, while far away  conductive propagation will rule 
the flame. The transition radius from one mode to the other, $R_\mathrm{tr}$, 
is a parameter of 
the problem which basically depends on the assumed thermal profile around the 
blob center at the time of ignition.

During the spontaneous combustion phase, the ignition timescale at a distance
$R$ from the center of a bubble is given by: 
\begin{equation}
\tau_{\mathrm i} \left[T\left(R\right)\right] = 
\tau_{\rm i}\left(T_{\mathrm b}\right) \exp\left[B\left(R/R_0\right)^2\right]\; 
.
\end{equation}
\noindent
The phase velocity of the flame, 
$v_\mathrm{sp} = \left(\D\tau_{\mathrm i}/\D R\right)^{-1}$, 
and the radius of a given bubble as a function 
of time can be easily obtained from the previous equation:
\begin{equation}
v_\mathrm{sp} \left(R\right) = 
\frac{R_0^2}{2B\tau_{\rm i}\left(T_{\mathrm b}\right)R} 
\exp\left[-B\left(\frac{R}{R_0}\right)^2\right]\; ,
\end{equation}
\begin{equation}
R_{\mathrm b} \left(t\right) = R_0 
\left\{\frac{1}{B}\ln\left[
\frac{t - \tau_{\rm i}\left(T_0\right) + \tau_{\rm i}\left(T_{\mathrm b}\right)}
{\tau_{\rm i}\left(T_{\mathrm b}\right)}\right]\right\}^{1/2}\; .
\end{equation}
\noindent
In the last equation, $\tau_{\rm i}\left(T_0\right)$ has to be evaluated at the
value of $T_0$ (central temperature of a hot spot at $t=0$) that makes it  possible that the
radius of the bubble be equal to $R_{\mathrm b}$ at a time $t$. This is easily
found to be:
\begin{equation}
T_0 \left(R_{\mathrm b}, t\right) = T_{\mathrm b} 
\left\{\frac{A}{t + \tau_{\rm i}\left(T_{\mathrm b}\right) - 
\tau_{\mathrm i} \left[T\left(R_{\mathrm b}\right)\right]}\right\}^{1/B}\; ,
\end{equation}
\noindent
where $\tau_{\mathrm i} \left[T\left(R_{\mathrm b}\right)\right]$ is given by 
Eq. (A.4). Finally, after taking the derivative of $T_0$ with respect
to $R_\mathrm{b}$, we arrive at 
the distribution function of the sizes of the bubbles during the spontaneous flame
phase, as a function of time $t$:
\begin{equation}
\left(\frac{\D N}{\D R_{\mathrm b}}\right)_\mathrm{sp} = 
\frac{2 T_{\mathrm b} 
\theta}{R_0^2} R_{\mathrm b}
\frac{\exp\left[B\left(\frac{R_\mathrm{b}}{R_0}\right)^2\right]}
{\left\{\frac{t}{A} + 1 - 
\exp\left[B\left(\frac{R_\mathrm{b}}{R_0}\right)^2\right]
\right\}^C}\; ,
\end{equation}
\noindent
with $\theta = \theta\left[T_0 \left(R_{\mathrm b}, t\right)\right]$, and 
$C = {\left(B+1\right)/B}$.

\begin{figure}
\resizebox{\hsize}{!}{\includegraphics{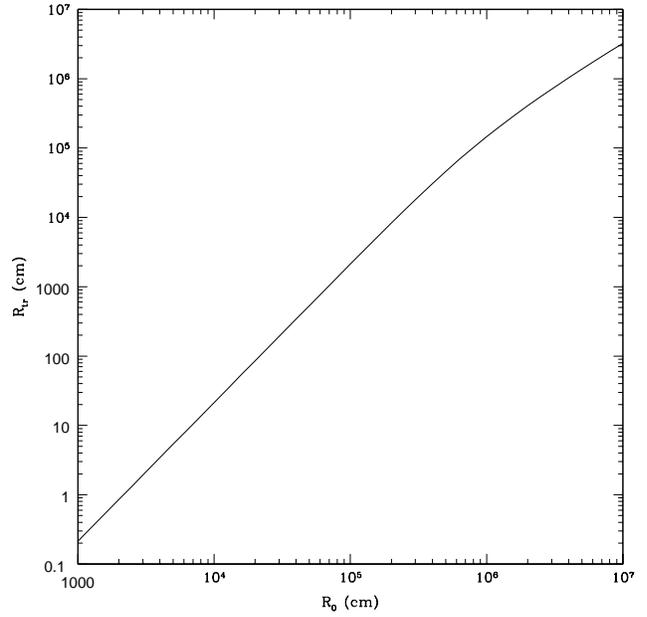}}
\caption{Radius of transition from spontaneous combustion to conductive flame
as a function of the characteristic lengthscale of the thermal profile of the 
bubbles.
}\label{figa1}
\end{figure}

The radius of transition from spontaneous combustion to conductive flame can be
calculated as the radius at which the phase velocity given by Eq. (A.5) equals
the laminar flame velocity, $v_\mathrm{cond} \approx
6.7\times10^6$~cm\,s$^{-1}$ at the densities of interest. The results of this
calculation can be seen in Fig.~\ref{figa1}.

In our simplified picture, once a bubble attains the transition radius, the
flame begins to propagate at the conductive velocity. As this conductive
velocity is independent of the initial thermal profile of the hot spot, the 
distribution function of the bubbles of size
$R_\mathrm{b}$ at time $t$ is the same as the distribution function at a size
equal to
the transition radius at the time $t - \Delta\left(R_{\mathrm b}\right)$, 
where $\Delta\left(R_{\mathrm b}\right) = \left(R_{\mathrm b} -
R_\mathrm{tr}\right)/v_\mathrm{cond}$, i.e.,
\begin{equation}
\left(\frac{\D N}{\D R_{\mathrm b}}\right)_\mathrm{co} 
\left(R_{\mathrm b}, t\right) = 
\left(\frac{\D N}{\D R_{\mathrm b}}\right)_\mathrm{sp}
\left[R_\mathrm{tr}, t - \Delta\left(R_{\mathrm b}\right)\right]\; .
\end{equation}
\noindent
The final results are shown in Figs.~\ref{fig3} and \ref{fig4}. 

We have tested the sensitivity of the results to different assumptions about 
the thermal profile within the bubbles at ignition (eq. A.3), but we have not
found any significant differences. The main 
limitations of our approach derive from the assumptions made in order to make 
the analytical calculation affordable. As discussed in more detail in Sect. 2 
we have not considered changes in density during combustion, neither  have we  
included heat transport by turbulence, which could 
influence the early phases of the ignition. During the first stage 
of bubble combustion following central ignition, dominated by spontaneous burning, there is no time for 
turbulence to contribute appreciably to the transport of thermal energy. Later on, conduction dominates but this 
is probably only true for small fluid elements where the ratio surface/volume 
is high. On the other hand, the statistical distribution of peak temperatures 
at t=0 s could be modified by turbulence before ignition. However, we have 
shown that the precise form of this statistical distribution is not relevant for the results of our analytical model.


\begin{thebibliography}{}

\bibitem[2004]{b04} Badenes, C. 2004, PhD Thesis, Universitat Polit\`ecnica de
Catalunya

\bibitem[2003]{blh03} Baron, E., Lentz, E.J., \& Hauschildt, P.H. 2003, ApJ, 588, 
L29

\bibitem[2003]{bgs03} Bravo, E., \& Garc\'\i a-Senz, D. 2003, in From Twilight 
to Highlight: The physics of supernovae, ed. W. Hillebrandt, \& B. Leibundgut
(Berlin: Springer), 165

\bibitem[2004]{bgs04} Bravo, E., \& Garc\'\i a-Senz, D. 2004, in Supernovae (10
years of SN1993J), eds. J.M. Marcaide, \& K.W. Weiler, in press.  

\bibitem[2004]{cgsb04} Cabezon, R.M., Garc\'\i a-Senz D, \& Bravo, E. 2004,
ApJS, 151, 345

\bibitem[2003]{gkochr03} Gamezo, V.N., Khokhlov, A.M., Oran, E.S., 
Chtchelkanova, A.Y., \& Rosenberg, R.O. 2003, Science, 299, 77 (G03)

\bibitem[2003]{gsb03} Garc\'\i a-Senz D., \& Bravo, E. 2003, in From Twilight to 
Highlight: The physics of supernovae, ed. W. Hillebrandt, \& B. Leibundgut
(Berlin: Springer), 158

\bibitem[1998]{gsbs98} Garc\'\i a-Senz D., Bravo, E., \& Serichol, N. 1998, ApJS, 
115, 119

\bibitem[1995]{gsw95} Garc\'\i a-Senz D., \& Woosley, S.E. 1995, ApJ, 454, 895

\bibitem[2000]{hblbfg00} Hatano, K., Branch, D., Lentz, E.J.,Baron, E., 
Filippenko, A.V., \& Garnavich, P.M. 2000, ApJ, 543, L49 

\bibitem[2002]{hs02} H\"oflich, P.A., \& Stein, J. 2002, ApJ, 568, 779

\bibitem[2003]{knw03} Kasen, D., Nugent, P., Wang, L., Howell, D.A., 
Wheeler, J.C., H\"oflich, P., Baade, D., Baron, E., \& Hauschildt, P.H. 2003, 
ApJ, 593, 788

\bibitem[1990]{k90} Khokhlov, A.M. 1990, MPA Report number 513

\bibitem[1995]{k95} Khokhlov, A.M. 1995, ApJ, 449, 695

\bibitem[2003]{mhvw03} Marion, H\"oflich, P., Vacca, W.D., \& Wheeler, J.C. 
2003, ApJ, 591, 316

\bibitem[2002]{n02} Napiwotzki, R. et al. (SPY consortium) 2002, A\&A, 386,
957

\bibitem[1996]{nhw96} Niemeyer, J., Hillebrandt, W., \& Woosley, S.E. 1996, ApJ,
471, 903

\bibitem[1997]{n97} Nugent, P., Baron, E., Hauschildt, P., \& Branch, D. 1997,
ApJ, 485, 812

\bibitem[2002]{rhn02} Reinecke, M., Hillebrandt, W., \& Niemeyer, J. 2002, A\&A, 
391, 1167 (RHN)

\bibitem[1998]{sn98} Saio, H., \& Nomoto, K. 1998, ApJ, 500, 388

\bibitem[1997]{scm97} Segretain, L., Chabrier, G., \& Mochkovitch, R. 1997, ApJ,
481, 355

\bibitem[2002]{tkbb02} Thomas, R.C., Kasen, D., Branch, D., \& Baron, E. 2002, 
ApJ, 567, 1037

\bibitem[2000]{thw00} Timmes, F.X., Hoffman, R.D., \& Woosley, S.E. 2000,
ApJS, 129, 377


\bibitem[1992]{tw92} Timmes, F.X., \& Woosley, S.E. 1992, ApJ, 396, 649

\bibitem[1992]{tw92} Woosley, S.E., Wunsch, S., \& Kuhlen, M.,  2004, ApJ, 607,
921 

\end{thebibliography}
\end{document}